\documentclass[aps,prx,twocolumn,superscriptaddress,showpacs]{revtex4-1}
\usepackage{amsmath,amssymb,graphics,epsfig,epstopdf,color,verbatim,ulem,braket,tabularx}
\usepackage{multirow}
\usepackage[colorlinks,linkcolor=blue,citecolor=blue,urlcolor=blue,bookmarks=false]{hyperref}
\usepackage{listings}

\begin{document}

\title{$(d-2)$-dimensional edge states of rotation symmetry protected topological states}

\author{Zhida Song}
\affiliation{Beijing National Laboratory for Condensed Matter Physics and Institute
of Physics, Chinese Academy of Sciences, Beijing 100190, China}
\affiliation{University of Chinese Academy of Sciences, Beijing 100049, China}

\author{Zhong Fang}
\affiliation{Beijing National Laboratory for Condensed Matter Physics and Institute
of Physics, Chinese Academy of Sciences, Beijing 100190, China}
\affiliation{Collaborative Innovation Center of Quantum Matter, Beijing, 100084,
China}

\author{Chen Fang}
\email{cfang@iphy.ac.cn}
\affiliation{Beijing National Laboratory for Condensed Matter Physics and Institute
of Physics, Chinese Academy of Sciences, Beijing 100190, China}

\begin{abstract}
We study fourfold rotation invariant gapped topological systems with time-reversal symmetry in two and three dimensions ($d=2,3$). We show that in both cases nontrivial topology is manifested by the presence of the $(d-2)$-dimensional edge states, existing at a point in 2D or along a line in 3D. For fermion systems without interaction, the bulk topological invariants are given in terms of the Wannier centers of filled bands, and can be readily calculated using a Fu-Kane-like formula when inversion symmetry is also present. The theory is extended to strongly interacting systems through explicit construction of microscopic models having robust $(d-2)$-dimensional edge states.
\end{abstract}
\maketitle

\textit{Introduction.}
A symmetry protected topological state (SPT) is a gapped quantum state that cannot be continuously deformed into a product state of local orbitals without symmetry breaking \cite{Gu2009,Chen2012,prb.86.125119}. SPT is known to have gapless boundary states in one lower dimension \cite{Chen2011a}, i. e., the $(d-1)$-dimensional edge, such as the spin-1/2 excitations at the end of a Haldane chain \cite{Haldane1983} or the Dirac surface states at the surface of a topological insulator \cite{Hasan2010,Qi2011}. The gapless states are protected by the symmetries on the $(d-1)$-dimensional edge, and when the symmetry is a spatial symmetry, they only appear on the boundary that is invariant under the symmetry operation \cite{Fu2011,Turner2010,Hughes2011,Hsieh2012}.

Very recently, the possibility of having gapped $d-1$-dimensional edge but gapless $d-2$-dimensional edge has been discussed \cite{Benalcazar2017,Schindler2017,codimension2,boundaryGF}. In Ref. \cite{Benalcazar2017}, it was shown that in a 2D spinless single particle (i.e., no spin-orbit coupling) system that has anti-commuting mirror planes, all four side edges can be gapped without symmetry breaking on an open square, but there are four modes localized at the four corners (0D edge) protected by mirror symmetries. Here we first extend the theory of 0D edge states to spin-1/2 fermion systems without mirror symmetries but with fourfold rotation symmetry and time-reversal symmetry. We point out that the presence of 0D-edge states can be understood as the result of a mismatch between the locations of the centers of the Wannier states and those of atoms. Then we generalize the theory to 3D, and define a new topological invariant by classifying the `spectral flow' of the Wannier centers between the $k_z=0$- and the $k_z=\pi$-slices in the Brillouin zone. When this invariant is nontrivial, there are four helical edge modes on the otherwise gapped side surfaces of the 3D system. We further show that when space inversion is also present, there is a Fu-Kane-like formula \cite{Fu2007} relating this invariant to certain combinations of rotation and inversion eigenvalues of the filled bands at high-symmetry crystal momenta. Finally, we generalize the theory to strongly interacting systems, by constructing microscopic models of boson and fermion SPT states that have $(d-2)$-dimensional edge states for $d=2,3$ using coupled wires construction. We remark that these edge states, protected by $C_4$ and some local symmetry such as time-reversal, are not pinned to the corners or hinges of the system, and can even appear in geometries having smooth side surfaces.

\begin{figure}
\begin{centering}
\includegraphics[width=1\linewidth]{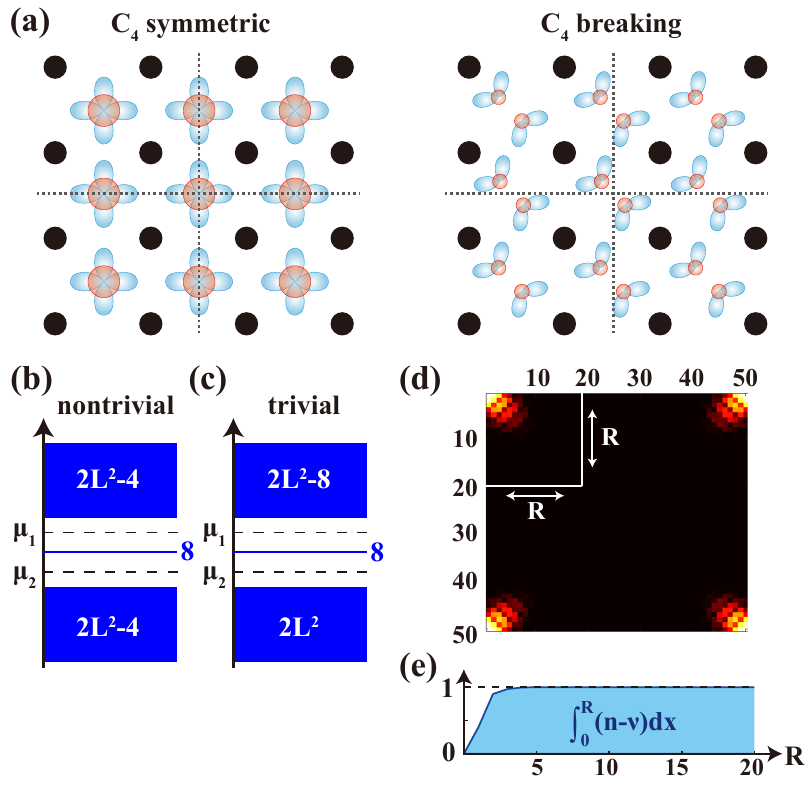}
\par\end{centering}
\protect\caption{\label{fig:2D}Nontrivial 0D edge modes of 2D fermion. In (a) we sketch the mismatch between the atom sites and the WCs in the presence (left panel)  and the absence (right panel) of $C_4$, where the atom sites are represented by the block circles and the WCs are represented by the colored orbitals. In (b) and (c), level counting for systems with nontrivial and with trivial 0D edge state are shown, respectively. In (d) the numerical calculated density profile of the 2D model with a finite size of $50\times50$ is plotted, where the Fermi level is set at $\mu_1$. The four bright regions in (d) show the additional charges located at the corners. To count the number of additional charges around a corner, we plot the integral of the density deviation from the filling ($\nu=4$) in (e).}
\end{figure}

\textit{Mismatch between the atom sites and the Wannier centers.}
Wannier functions for the filled bands can be constructed for all 2D gapped insulators that have zero Chern number \cite{Marzari1997}. When symmetries are involved (time-reversal and/or spatial), the set of Wannier functions may or may not form a representation of the symmetry group \cite{Soluyanov2011}. If they do, then we call these Wannier functions `symmetric'. If a set of symmetric Wannier functions cannot be found for all filled bands, we know that the system cannot be adiabatically deformed into an atomic insulator: this is considered a generalized definition of topologically nontrivial insulators \cite{Po2017,Bradlyn2017}, since atomic orbitals automatically form a set of symmetric wavefunctions. Atomic insulators are usually considered trivial. Nevertheless, we realize that even they can also be somewhat nontrivial if there is a mismatch between the Wannier centers and the atomic positions, as shown in the left panel of Fig. \ref{fig:2D}(a). A Wannier center (WC) can be understood as the middle of the Wannier function (but see Ref. \cite{SupMat} for a rigorous definition), and if the Wannier functions are symmetric, their centers are also symmetric. When the mismatch happens, it means that while the insulator can be deformed into some atomic insulator, it would \emph{not} be made by the atoms forming the lattice. The presence of 0D edge states of the system put on an open disk is the manifestation of the `mismatch'.

To be specific, let us consider a square lattice model
\begin{align}
H & =\left(1-\cos k_{x}-\cos k_{y}\right)\tau_{0}\sigma_{z}s_{0}+\sin k_{x}\tau_{0}\sigma_{x}s_{x}\nonumber \\
 & +\sin k_{y}\tau_{0}\sigma_{x}s_{y}+\Delta\left(\cos k_{x}-\cos k_{y}\right)\tau_{y}\sigma_{y}s_{0},\label{eq:H-2D}
\end{align}
in which all the atomic orbitals are put on the lattice sites. Here $\tau_{i}$, $\sigma_{i}$ ($i=0,x,y,z$) are Pauli matrices representing the orbital degrees of freedom, and $s_{i}$ ($i=0,x,y,z$) representing the spin. This model can be thought as two copies of 2D topological insulator plus a mixing term with $\Delta$ as coefficient; and it has time-reversal symmetry $T=-is_{y}K$ and a rotation symmetry $C_{4}=\tau_{z}e^{-i\pi s_{z}/4}$. The system put on a torus is fully gapped because the four terms in Eq. (\ref{eq:H-2D}) anti-commute with each other and their coefficients do not vanish at the same time.

As shown in Ref. \cite{SupMat}, whatever value $\Delta$ takes, the insulator is equivalent to an atomic one, and its WCs are located at the plaquette centers. We have explicitly constructed a set of symmetric Wannier functions and prove that, protected by the time-reversal and $C_4$ symmetries, the Wannier centers stay invariant under any gauge transformation that keeps the Wannier functions symmetric. This model hence realizes the mismatch between the WCs at plaquette centers and the atomic positions at sites.

Now we cut along the dotted lines in the left panel of Fig. \ref{fig:2D}(a) and turn the 2D torus into an open square. Since this cut preserves $C_{4}$ symmetries, the states centered at the plaquette center will be equally divided into the four quarters, so that each quarter carries one extra electron on top of some even integer filling. Due to $T$, this means that a pair (Kramers' pair) of zero modes are located near each of the four corners of the square. One may observe that in the absence of particle-hole symmetry (which is an accidental symmetry of the model), the modes can be moved away from zero and pushed into the bulk states, but we argue that even when this happens, the corners are still nontrivial in the following sense. The total eight modes (two near each corner) come from both the conduction and the valence bands, each having $(N_{band}-\nu){L^2}-4$ and $\nu{L}^2-4$ electrons respectively, where $N_{band}\in{even}$ and $\nu\in{even}$ are the total number of bands and the filling number respectively, and $L$ the length of the square (Fig. \ref{fig:2D}(b)). No matter where the Fermi energy is, a gapped ground state must have $4\ \textrm{mod}\ 8$ electrons on an even-by-even lattice, so that each corner has exactly one (or minus one) extra electron on top of the filling of the bulk. This is in sharp contrast with the systems having trivial corner states, whose energy levels are plotted in Fig. \ref{fig:2D}(c). In that case, the in-gap states can be pushed into the conduction bulk and there is no extra charge at each corner. In Fig. \ref{fig:2D}(d)-(e), we plot the charge density at $\mu=\mu_1$ in real space, and plot the extra electric charge within a small area near the corner as a function of radius in the Slater-product many-body ground state.

To see how the odd parity of the corner charge is protected by $C_4$, we contrast the above scenario with the one having a nematic perturbation breaking $C_4$ down to $C_2$, so that the Wannier centers are shifted to the positions shown in the right panel of Fig.\ref{fig:2D}(a). When the system is cut along the dotted lines, quarter has inside it an integer number of Kramers' pairs, and the degeneracy at each corner is absent.

\begin{figure}
\begin{centering}
\includegraphics[width=1\linewidth]{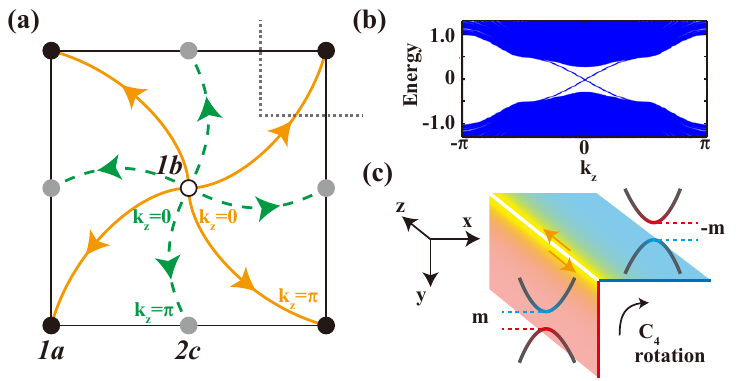}
\par\end{centering}
\protect\caption{\label{fig:3D}Nontrivial 1D helical modes of 3D insulator. In (a) we plot the two generators of nontrivial $\mathbb{Z}_2$-flows from the $k_z=0$-slice to the $k_z=\pi$-slice, where the lattice site ($1a$), the plaquette center ($1b$), and the edge midpoint ($2c$) are represented by black planchet, hollow circle, and grey planchet, respectively. In (b) the numerically calculated helical modes of our 3D model on a tetragonal cylinder geometry is plotted. The length along $x$ and $y$ directions is 50. In (c) we sketch the domain wall between surfaces of opposite masses, enforced by the $C_4$ rotation symmetry.}
\end{figure}

\textit{1D helical state and $\mathbb{Z}_2$ Wannier center flow.}
A natural generalization of the 0D state in 2D is the 1D edge state in 3D, where both the 3D bulk and 2D side surfaces are insulating, as shown in Fig. \ref{fig:3D}(c). Our construction of this state is also based on the WC picture. Assume the 3D system has $T$ and $C_{4}$ symmetries, so we can take a $C_{4}$ invariant tetragonal cell and transform the Hamiltonian along the $z$-direction to momentum space. Each slice with fixed $k_{z}$ can be thought as a 2D system, wherein the $k_{z}=0,\pi$-slices are time-reversal and $C_{4}$ invariant while the others are only $C_{4}$ invariant. Consider an insulator that has four filled bands, or four WCs for each $k_z$-slice. Due to $C_4$, the four WCs are related to each other by fourfold rotations; and due to $T$, at $k_z=0$ or $k_z=\pi$, two WCs that form a Kramers' pair must coincide. Therefore, at $k_z=0$ and $k_z=\pi$, there are only three possible configurations for the four WC: all four at $1a$, all four at $1b$ and two at each $2c$ Wyckoff positions. Wyckoff positions are points in a lattice that are invariant under a subgroup of the lattice space group. For a square lattice in a Wigner-Seitz unit cell, $1a$ and $1b$ are the center and the corner invariant under $C_4$, $2c$ are the middles of the edges invariant under $C_2$, and 4d are generic points invariant under identity (the trivial subgroup). If the configurations at $k_z=0$ and at $k_z=\pi$ are different, the evolution of the WC between the two slices forms a `$\mathbb{Z}_2$-flow', a robust topological structure revealing that the 3D insulator is not an atomic one. Out of several different combinations of the configurations at $k_z=0$ and $k_z=\pi$, there are two topologically distinct $\mathbb{Z}_2$-flows, where the four WCs flow from $1b$ to $1a$ and from $1b$ to $2c$ [solid yellow and dashed green lines in Fig. \ref{fig:3D}(a)], respectively. The latter $\mathbb{Z}_2$-flow can be shown equivalent to a weak topological index (Ref. \cite{SupMat}), and we from now on focus on the first $\mathbb{Z}_2$-flow from $1b$ to $1a$. Whether this flow is present or not gives us a new $\mathbb{Z}_2$-invariant, and its edge manifestation is the existence of 1D helical edge modes on the side surface of a bulk sample. (For more rigorous definition and classification of the WC flow for arbitrary number of filled bands, see Ref. \cite{SupMat}.)

To see this bulk-edge correspondence, we cut the bulk along both $x$ and $y$ directions, keeping the periodic boundary condition along $z$. From top-down perspective, a corner of the sample takes the shape of the dotted lines shown in Fig. \ref{fig:3D}(a). One can see that at the corner, the boundary cuts through exactly one (or three) line in the WC flow, corresponding to one helical mode along the hinge between the two open surfaces. To make the picture more concrete, we consider the following 3D model, which is a simple extension of the 2D model in Eq. (\ref{eq:H-2D}).
\begin{align}
H & =\left(2-\sum_{i}\cos k_{i}\right)\tau_{0}\sigma_{z}s_{0}+\sum_{i}\sin k_{i}\tau_{0}\sigma_{x}s_{i}\nonumber \\
 & +\Delta\left(\cos k_{x}-\cos k_{y}\right)\tau_{y}\sigma_{y}s_{0}\label{eq:H-3D}
\end{align}
The $k_{z}=0$-slice is equivalent with the 2D model in Eq. (\ref{eq:H-2D}), thus having four charges locating at the plaquette center. The $k_{z}=\pi$-slice is, however, a 2D atomic insulator with four charges locating at the lattice site. The mismatch between the WCs at $k_z=0$- and $k_z=\pi$-slices means that the $\mathbb{Z}_2$-flow exists.
To confirm the $\mathbb{Z}_{2}$-flow, we also choose a smooth gauge for all the $k_{z}$ slices from $k_{z}=0$ to $k_{z}=\pi$ and plot the WC flow explicitly, which indeed gives the $\mathbb{Z}_2$-flow, shown in Ref. \cite{SupMat}.
The 1D helical state is also confirmed by a numerical calculation of the band structure of a finite tetragonal cylinder, as plotted in Fig. \ref{fig:3D}(b). For this particular model, the helical edge states can be viewed from another perspective. The edge between the two open surfaces can be considered as the domain wall between them. On each surface there is a mass gap, and the rotation symmetry in this model enforces the two masses to be opposite, so that at the domain wall there is a helical mode \cite{Hsieh2012} [see Fig. \ref{fig:3D}(c) for a schematic and see Ref. \cite{SupMat} for more details].

\textit{Symmetry indicators for the $\mathbb{Z}_2$-invariant.}
To see if a given insulator has 1D helical edge modes on the side surface, one needs to calculate the evolution of the WCs as a function of $k_z$, which in turn requires finding symmetric, smooth and periodic Bloch wave functions for all bands at each $k_z$-slice as is done for our model Hamiltonian. This is practically impossible in real materials. Now we show that in the presence of additional inversion symmetry, this $\mathbb{Z}_2$-invariant can be determined by the rotation and inversion eigenvalues at all high-symmetry momenta, simplifying the diagnosis. We call this method a ``Fu-Kane-like formula'',  likening it to the Fu-Kane formula for time-reversal topological insulators \cite{Fu2007}, where inversion is not required to protect the nontrivial topology, but when present greatly simplifies the calculation.

This formula is derived based on the new theory of symmetry indicators \cite{Po2017,Bradlyn2017}: given any insulator, a full set of eigenvalues of the space group symmetry operators for filled bands at all high-symmetry points generates a series of indicators. They tell us if this set is consistent with any atomic insulator, and if yes, the theory further gives where the atomic orbitals are located. Our goal is to find such an indicator that is equivalent to the $\mathbb{Z}_2$-invariant for the WC flow. Following the WC flow picture, we require: (i) at $k_z=0$ and $k_z=\pi$, the eigenvalues of $C_4$, $C_2=C_4^2$ and $P$ are consistent with atomic insulators; (ii) there is no surface state on the side surfaces; and (iii) comparing the two slices at $k_z=0$ and $k_z=\pi$, the numbers of atomic orbitals at $1a$ and at $1b$ change by $\pm4$ and $\mp4$, respectively. For a concrete example, let us consider space group $P4/m$, whose indicators form a group $\mathbb{Z}_2\times\mathbb{Z}_4\times\mathbb{Z}_8$ \cite{Po2017}, so that insulator according to its $C_4$ and $P$ eigenvalues can be denoted by $(mnl)$ ($m=0,1$, $n=0,1,2,3$, $l=0,1,...,7$), and an insulator with a nonzero indicator cannot be adiabatically deformed into an atomic insulator. Using the three criteria above, we find that the $\mathbb{Z}_2$-flow is nontrivial only if $(mnl)=(004)$. We have found the explicit formulas to calculate these indicators directly from the symmetry eigenvalues, which can be applied to all space groups having both C4 and P. (See Ref. \cite{SupMat} for the results, and find a MATLAB script therein for automated diagnosis for materials in these space groups.)

\textit{Extension to strongly interacting SPT.}
In the above we have established the theory of $(d-2)$-dimensional edge modes for free fermions through the WC picture. Since WC is a single particle object, the same picture does not apply for strongly interacting bosons or fermions. Here we rebuild a 3D free fermion model with robust 1D helical edge modes using coupled wires construction \cite{Teo2014,Oreg2014,Neupert2014,wire-QSH,wire-QAH}, a method that can be easily extended to strongly interacting SPT. These SPT can either be bosonic \cite{Thorngren2016} or fermionic, and are in general protected by spatial symmetry \cite{Huang2017} plus some internal symmetry \cite{Wang2017}.

\begin{figure}
\begin{centering}
\includegraphics[width=1\linewidth]{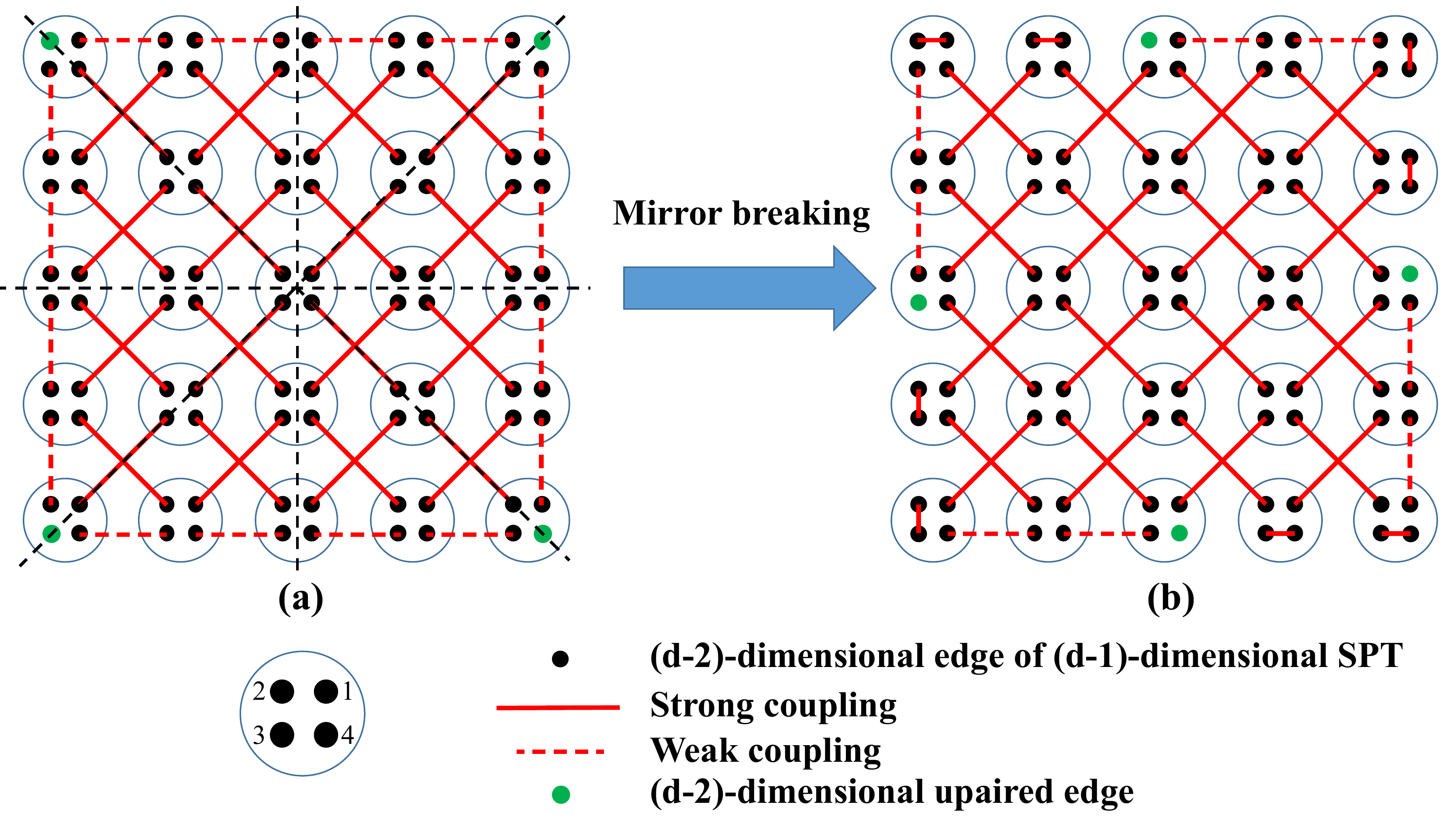}
\par\end{centering}
\protect\caption{\label{fig:3}Coupled wires construction for a 3D SPT with robust 1D edge modes. Each filled circle is a wire from topdown, and each open circle including four elementary wires is a `physical' wire that can be realized in 1D lattice models. The breaking of mirror symmetry in (a) causes the edge modes to move to the positions in (b).}
\end{figure}

Consider an arrangement of 1D wires shown in topdown view in Fig. \ref{fig:3}(a), each of which represents a helical mode. Due to the fermion doubling theorem, each wire alone cannot be physically realized in 1D, but an even number of these wires can be realized as a 1D wire fine tuned to a critical point. In our model, four wires make a physical, critical 1D wire. For concreteness we assume that under $C_4$-rotation the four wires inside cyclically permute. Then we couple the wires in the following way: the four wires, in topdown view, which share a plaquette are coupled diagonally, i.e., 1 coupled to 3 and 2 to 4. For a 3D torus these couplings (solid red lines) make the coupled wire system an insulator. For a cylinder geometry open in $x$ and $y$ directions, however, there are `dangling helical wires' on the side surfaces, which can again be gapped by turning on a dimerizing coupling (dotted lines). But one soon discovers that, as long as $C_4$ is preserved, there are always four unpaired wires on the side surface (represented by green dots), which are in fact the same 1D helical edge mode protected by $C_4$ and $T$ studied above.

This construction can be easily extended to strongly interacting SPT. One simply replaces each helical wire with a $(d-2)$-dimensional edge of a $(d-1)$-dimensional SPT protected by some local symmetry. For example, each `wire' can be a 0D spin-1/2, which is the edge of a 1D Haldane chain protected by SO(3) symmetry. In that case, the resultant construction in Fig. \ref{fig:3}(a) is nothing but an AKLT-like state \cite{Afflect1987,Affleck1988} formed by $S=2$ spins, but unlike previously considered AKLT states in 2D, it has gapped 1D edge but four 0D gapless spin-1/2 excitations localized at the four corners in an open square. We can also replace each wire by the edge of a Levin-Gu state \cite{Levin2012}, protected by a $\mathbb{Z}_2$ local symmetry, then the construction in Fig. \ref{fig:3}(a) is a 3D bosonic SPT with 1D gapless modes at four corners. Notice that in these boson examples, time-reversal symmetry is not necessary. Similar construction can be used to obtain SPT states protected by both the local symmetries (being $T$, SO(3) or $Z_2$) and $C_4$-rotation symmetry.

\textit{Discussion.}
It is important to note that, while in examples studied so far, the $(d-2)$-dimensional edge modes sit at the corners or hinges in the disk or cylinder geometry, it is not always the case. In the model shown in Fig. \ref{fig:3}(a), the edge modes are pinned to the corners by the mirror symmetries (dotted lines), and breaking these mirror planes in the bulk or on the surface causes the edge modes to move away. In the example shown in Fig. \ref{fig:3}(b), we break the mirror symmetry of the construction on the surface, so that the dangling wires move from the corners to some generic points on the side. As long as $C_4$ is present, the $(d-2)$-dimensional edge modes are stable, yet not pinned to corners or hinges in the absence of mirror symmetries. In fact, they still appear even if the whole side surface is smooth without hinges at all. We also emphasize that while these edge modes are protected by $C_4$-rotation symmetry, breaking the symmetry perturbatively in the bulk or on the boundary does not in general gap out the modes, because time-reversal alone is sufficient to protect 1D helical edge modes. The only way of gapping the modes is to annihilate them in pairs, and this means large $C_4$-breaking either in the bulk or on the boundary. Similar discussions may be extended to systems with twofold, threefold, sixfold rotations.

Experimentally, the four helical edge modes of a 3D electronic insulator contribute a quantized conductance of $4e^2/h$ that may be measured in electric transport \cite{Qi2011}. Also the $(d-2)$-dimensional edge modes may be detected by local probes such as scanning tunneling microscopy, either on a bulk sample or at the step edge of a thin film.

\textit{Acknowledgement.}
C. F. thanks Xi Dai, Meng Cheng, Yang Qi and B. Andrei Bernevig for helpful discussion. The work was supported by the National Key Research and Development Program of China under grant No. 2016YFA0302400, and by NSFC under grant No. 11674370.

\textit{Note added.}
We are aware of works on related topics that have appeared on arXiv after our posting \cite{Hughes-d-2,Titus-d-2,Langbehn-d-2}; their results have finite overlap with ours and seem consistent.

\bibliography{TI}

\pagebreak

\setcounter{equation}{0}
\setcounter{figure}{0}
\setcounter{table}{0}
\setcounter{page}{1}

\makeatletter
\renewcommand{\theequation}{S\arabic{equation}}
\renewcommand{\thefigure}{S\arabic{figure}}
\renewcommand{\thetable}{S\Roman{table}}

\definecolor{mygreen}{RGB}{28,172,0} 
\definecolor{mylilas}{RGB}{170,55,241}

\lstset{language=Matlab,%
    breaklines=true,%
    morekeywords={matlab2tikz},
    keywordstyle=\color{blue},%
    morekeywords=[2]{1}, keywordstyle=[2]{\color{black}},
    identifierstyle=\color{black},%
    stringstyle=\color{mylilas},
    commentstyle=\color{mygreen},%
    showstringspaces=false,
    numbers=left,%
    numberstyle={\tiny \color{black}},
    numbersep=9pt, 
}

\onecolumngrid
\begin{center}
\textbf{ \large Supplemental materials for ``$(d-2)$-dimensional edge states of rotation symmetry protected topological states''}
\end{center}
\bigskip
\twocolumngrid

\section{A brief review of Wannier functions} \label{sub:WF}

Wannier functions (WFs) are defined as the Fourier transformations of Bloch wave functions,
with a gauge freedom
\begin{equation}
\left|w_{\alpha\mathbf{R}}\right\rangle =\frac{1}{\sqrt{\mathcal{N}}}\sum_{\mathbf{k}\alpha}e^{-i\mathbf{k}\cdot\mathbf{R}}\left|\phi_{\alpha\mathbf{k}}\right\rangle
\end{equation}
\begin{equation}
\left|\phi_{\alpha\mathbf{k}}\right\rangle =\sum_{n}U_{n\alpha}\left(\mathbf{k}\right)\left|\psi_{n\mathbf{k}}\right\rangle
\end{equation}
Here $\left|\psi_{n\mathbf{k}}\right\rangle $ is the $n$-th Bloch
state at $\mathbf{k}$, $\mathbf{R}$ is a lattice vector in real
space, $\mathcal{N}$ is the number of cells, $\alpha$ is the wannier
index, and $U\left(\mathbf{k}\right)$ can be an arbitray unitary
matrix.
Generally speaking, as long as $\left|\phi_{\alpha\mathbf{k}}\right\rangle $
takes a smooth gauge in the whole Brillouin zone, the corresponding WF $|w_{\alpha \mathbf{R}}\rangle$ will be well localized around the lattice $\mathbf{R}$.

The WC (Wannier center) is defined as the position expectation on the WF.
It can be calculated from the Berry connection in momentum space
\begin{equation}
\left\langle w_{\alpha\mathbf{R}}\right|\hat{\mathbf{x}}\left|w_{\alpha\mathbf{R}}\right\rangle =\int \frac{d^{d}\mathbf{k}}{(2\pi)^d}\ \boldsymbol{\mathcal{A}}_{\alpha}\left(\mathbf{k}\right)+\mathbf{R}
\end{equation}
where the Berry connection is defined as
\begin{equation}
\boldsymbol{\mathcal{A}}_{\alpha}\left(\mathbf{k}\right)=i\sum_{mn}\left\langle U_{m\alpha}u_{m}\right|\partial_{\mathbf{k}}\left|U_{n\alpha}u_{n}\right\rangle
\end{equation}
and $u_{n}\left(\mathbf{k}\right)$ is the periodic part of the
Bloch wavefunction $\left|\psi_{n\mathbf{k}}\right\rangle $.
In general, WCs are gauge dependent, i.e. they depend on the choice of the gauge $U_{n\alpha}(\mathbf{k})$.
An exception is the 1D insulating system where the 1D WCs can be thought as the spectrum of
 the Wilson loop along the 1D Brillouin zone and thus are gauge invariant quantities.
Such a property leads to the modern theory of polarization
and has greatly facilitated the studies of topological insulators because the spectrum of
 wilson loop is believed to be isomorphic with the surface dispersion.

To define the symmetric WFs, let us firstly review the concept
of Wyckoff positions. General positions in the unit
cell can be classified into a few types of Wyckoff positions by the
their site symmetry groups (SSGs). The SSG for a given site $\mathbf{x}_{1}$
can be defined as the collection of all the space group elements that
leave $\mathbf{x}_{1}$ invariant (module a lattice vector), i.e.
\begin{equation}
G\left(\mathbf{x}_{1}\right)=\left\{ g\in G|\exists\mathbf{R}\ s.t.\ g\mathbf{x}_{1}=\mathbf{x}_{1}+\mathbf{R}\right\}
\end{equation}
Here $g\mathbf{x}=p_{g}\mathbf{x}_{1}+\mathbf{t}_{g}$ with $p_{g}$
the point group operation of $g$ and $\mathbf{t}_{g}$ the translational
operation of $g$.
The equivalent positions of $\mathbf{x}_{1}$ can be generated
from a complete set of the representatives of the quotient group $G/G\left(\mathbf{x}_{1}\right)$
\begin{equation}
\mathbf{x}_{\sigma}=g_\sigma \mathbf{x}_{1}-\left[g_\sigma \mathbf{x}_{1}\right]
\qquad
g_\sigma \in G/G(\mathbf{x}_1)
\end{equation}
where $g_{1}=e$ is the identity, and $\left[g_\sigma\mathbf{x}_{1}\right]$
is the lattice containing $g_\sigma\mathbf{x}_{1}$ such that $\mathbf{x}_{\sigma}$
locates inp the home cell. Hereafter, for convenience some times we
will split the Wannier index $\alpha$ into a site index $\sigma$
and an orbital index $\mu$, indicating that the WF is the $\mu$-th
orbital locating at the site $\mathbf{x}_{\sigma}$. A set of WFs
is called symmetric if (i) they form representations (reps) of the SSGs
\begin{equation}
\forall g\in G\left(\mathbf{x}_{\sigma}\right)\quad g\left|w_{\sigma\mu\mathbf{R}}\right\rangle =\sum_{\nu}D_{\nu\mu}\left(g\right)\left|w_{\sigma\nu,\mathbf{R}+[g\mathbf{x}_\sigma]}\right\rangle
\end{equation}
, (ii) WFs at a general equivalent position of $\mathbf{x}_{1}$ can
be generated from the WFs at $\mathbf{x}_{1}$ by a symmetry operation relating
the two positions, and (iii) for time-reversal invariant
systems, we further ask the WFs to form Kramers' pairs, i.e.
\begin{equation}
T\left|w_{\alpha\mathbf{R}}\right\rangle =\sum_{\beta}\Omega_{\beta,\alpha}\left|w_{\beta\mathbf{R}}\right\rangle
\end{equation}
where $\Omega$ is an anti-symmetric unitary matrix. For convenience in the following we take its standard form as
\begin{equation}
\Omega=\begin{bmatrix}0 & -\mathbb{I}\\
\mathbb{I} & 0
\end{bmatrix}\label{eq:omg}
\end{equation}
The transformations of $\left|\phi_{\sigma\mu\mathbf{k}}\right\rangle $
under symmetry operations are completely determined by the symmetry
property of WFs. For the space groups considered in our work, we have
\begin{equation}
\forall g\in G\qquad g\left|\phi_{\sigma\mu\mathbf{k}}\right\rangle =\sum_{\sigma^{\prime}\mu^{\prime}}D_{\sigma^{\prime}\mu^{\prime},\sigma\mu}^{\mathbf{k}}\left(g\right)\left|\phi_{\sigma^{\prime}\mu^{\prime}g\mathbf{k}}\right\rangle \label{eq:g-phi}
\end{equation}
\begin{equation}
D_{\sigma^{\prime}\mu^{\prime},\sigma\mu}^{\mathbf{k}}\left(g\right)=\delta_{\mathbf{x}_{\sigma^{\prime}},g\mathbf{x}_{\sigma}}^{\prime}e^{ig\mathbf{k}\cdot\left(\mathbf{x}_{\sigma^{\prime}}-g\mathbf{x}_{\sigma}\right)}D_{\mu^{\prime}\mu}\left(g\right)\label{eq:Dk-def}
\end{equation}
, where $\delta_{\mathbf{x}_{\sigma^{\prime}},g\mathbf{x}_{\sigma}}^{\prime}=1$
if $\mathbf{x}_{\sigma^{\prime}}=g\mathbf{x}_{\sigma}$ module
a lattice, and
\begin{equation}
T\left|\phi_{\alpha\mathbf{k}}\right\rangle =\sum_{\beta}\Omega_{\beta,\alpha}\left|\phi_{\beta,-\mathbf{k}}\right\rangle \label{eq:T-phi}
\end{equation}
The transformation matrices $D^{\mathbf{k}}$ and $\Omega$ will be referred as the sewing matrices
 in the following.

\section{Gauge invariant 2D Wannier centers}

The discussion about the mismatch between atom sites and WCs
in the text presumes the gauge invariance of the occupied 2D WCs.
Otherwise, we can choose a gauge where the WCs move away from
the plaquette center to negative all the arguments.
Similarly, the gauge invariance of the $\mathbb{Z}_2$-flow discussed
in the text also needs the 2D WCs at $k_{z}=0,\pi$-slices
to be gauge invariant. Here, we will set such a cornerstone by proving
 that in some special cases the 2D WCs indeed are gauge invariant
 guaranteed by the crystalline symmetry.

Since all the symmetry properties of symmetric WFs are encoded in the sewing matrices, in the proof below we follow the logic that two sets of symmetric WFs
 can be deformed to each other \emph{only if} the sewing matrices
 generated from them can be transformed to each other by a smooth gauge
 transformation.
A relevant useful concept is the band representation (BR), i.e. the set of irreducible reps (irreps) at high-symmetry momenta,
 which is indeed the diagonal blocks of sewing matrices in momentum space.
In some cases, the information in BR is enough to demonstrate two sets of WFs
 are inequivalent.
Therefore, before the proof, let us figure out what BR can tell us.
The smallest 2D space group containing $C_{4}$ is $p4$.
As shown in table \ref{tab:wkf-p4}, it has four types of wyckoff positions in real
space, wherein the $1a$ and $1b$ positions are the site and plaquette
center, respectively.
To describe the BR we only need to count the irreps at $\Gamma$ and $M$,
 because the irreps at $X$ and general momentum are always same.
After a few derivations according to Eq. (\ref{eq:Dk-def}), we find the mappings from symmetric WFs to BRs as
\begin{equation}
E_{\frac{1}{2}}^{1a}\mapsto E_{\frac{1}{2}}^{\Gamma}+E_{\frac{1}{2}}^{M}\label{eq:2Dmap-1a12}
\end{equation}
\begin{equation}
E_{\frac{3}{2}}^{1a}\mapsto E_{\frac{3}{2}}^{\Gamma}+E_{\frac{3}{2}}^{M}\label{eq:2Dmap-1a32}
\end{equation}
\begin{equation}
E_{\frac{1}{2}}^{1b}\mapsto E_{\frac{1}{2}}^{\Gamma}+E_{\frac{3}{2}}^{M}
\end{equation}
\begin{equation}
E_{\frac{3}{2}}^{1b}\mapsto E_{\frac{3}{2}}^{\Gamma}+E_{\frac{1}{2}}^{M}\label{eq:2Dmap-1b32}
\end{equation}
\begin{equation}
E_{\frac{1}{2}}^{2c}\mapsto E_{\frac{1}{2}}^{\Gamma}+E_{\frac{3}{2}}^{\Gamma}+E_{\frac{1}{2}}^{M}+E_{\frac{3}{2}}^{M}
\end{equation}
\begin{equation}
E_{\frac{1}{2}}^{4d}\mapsto2E_{\frac{1}{2}}^{\Gamma}+2E_{\frac{3}{2}}^{\Gamma}+2E_{\frac{1}{2}}^{M}+2E_{\frac{3}{2}}^{M}\label{eq:2Dmap-4d12}
\end{equation}
Here we use the symbol of an irrep decorated with a Wyckoff position
 to represent the WFs forming this irrep at this Wyckoff position,
 and use the symbol of an irrep decorated with a momentum to represent
 the bands forming this irrep at this momentum.
We follow the notations for irreps in Ref. [\onlinecite{point-group}].
In the following, we will make use of these mappings.

\begin{table}
\begin{centering}
\begin{tabular}{|c|c|c|c|c|c|}
\hline
SSG & W & K & Coordinates & irreps at W & irreps at K\tabularnewline
\hline
\hline
\multirow{2}{*}{$C_{4}$} & $1a$ & $\Gamma$ & $\left(0,0\right)$ & $E_{\frac{1}{2}}$ $E_{\frac{3}{2}}$ & $E_{\frac{1}{2}}$ $E_{\frac{3}{2}}$\tabularnewline
\cline{2-6}
 & $1b$ & $M$ & $\left(\frac{1}{2},\frac{1}{2}\right)$ & $E_{\frac{1}{2}}$ $E_{\frac{3}{2}}$ & $E_{\frac{1}{2}}$ $E_{\frac{3}{2}}$\tabularnewline
\hline
$C_{2}$ & $2c$ & $X$ & $\left(0,\frac{1}{2}\right)$ & $E_{\frac{1}{2}}$ & $E_{\frac{1}{2}}$\tabularnewline
\hline
$C_{1}$ & $4d$ & ... & $\left(x,y\right)$ & $E_{\frac{1}{2}}$ & $E$ \tabularnewline
\hline
\end{tabular}
\par\end{centering}

\raggedright{}\protect\caption{\label{tab:wkf-p4}The Wyckoff positions and high-symmetry momenta in 2D space group $p4$ (with time-reversal symmetry).
Due to the time-reversal symmetry, all the irreps at
Wyckoff positions and time-reversal invariant momenta are double degenerate.
}
\end{table}

\subsection{Gauge invariant Wannier centers at $1a$ (site)\label{sub:2DWF-1a}}

As will be proved latter, a set of WFs at $1a$ can be moved away by a symmetric
gauge transformation \emph{only if} they form a rep consist
of an even number of combined rep $E_{\frac{1}{2}}^{1a}+E_{\frac{3}{2}}^{1a}$.
Therefore, the conclusion is that, for a given set of WFs at $1a$, taking
off all the even number of the rep $E_{\frac{1}{2}}^{1a}+E_{\frac{3}{2}}^{1a}$,
the left WFs consisted of
\begin{equation}
n\left(E_{\frac{1}{2}}^{1a}+E_{\frac{3}{2}}^{1a}\right)+mE_{\frac{1}{2}}^{1a}+m^{\prime}E_{\frac{3}{2}}^{1a}
\end{equation}
, where $n=0,1$, and one of $m$ $m^{\prime}$ equals to zero, will
stay still under any symmetry allowed gauge transformation. We will
prove this statement in three steps.

Firstly, we will show that the eight WFs forming the
 rep  $2E_{\frac{1}{2}}^{1a}+2E_{\frac{3}{2}}^{1a}$
 can be moved to four Kramers' pairs at $4d$ positions without
 breaking any symmetry.
Apply an unitary transform for the bases in $2E_{\frac{1}{2}}^{1a}+2E_{\frac{3}{2}}^{1a}$
\begin{align}
\left|w_{1}\right\rangle & =e^{-i\frac{\pi}{4}}\left|\frac{1}{2}\right\rangle_A
   + e^{-i\frac{\pi}{4}}\left|\bar{\frac{1}{2}}\right\rangle_B \nonumber \\
 & + e^{i\frac{\pi}{4}}\left|\frac{3}{2}\right\rangle_A +
     e^{i\frac{\pi}{4}}\left|\bar{\frac{3}{2}}\right\rangle_B
\end{align}
\begin{equation}
\left|w_{i+1}\right\rangle = C_4 \left|w_{i}\right\rangle \qquad i=0,1,2
\end{equation}
\begin{equation}
\left|w_{i+4}\right\rangle = T \left|w_{i}\right\rangle \qquad i=0,1,2,3
\end{equation}
, where the subscript $A/B$ is used to distinguish the two same irreps,
we find that the time-reversal rep has the standard form $\Omega$, while the $C_{4}$ rep is
 identical with the $E_\frac{1}{2}^{4d}$ WFs
\begin{equation}
D\left(C_{4}\right)=\sigma_{0}\otimes\begin{bmatrix}0 & 0 & 0 & -1\\
1 & 0 & 0 & 0\\
0 & 1 & 0 & 0\\
0 & 0 & 1 & 0
\end{bmatrix}
\end{equation}
Therefore, the WFs in $2E_{\frac{1}{2}}^{1a}+2E_{\frac{3}{2}}^{1a}$ can be moved
 to $4d$ without breaking any symmetry.
We denote this equivalent relation as $2E_{\frac{1}{2}}^{1a}+2E_{\frac{3}{1}}^{1a}\sim E_{\frac{1}{2}}^{4d}$.
Readers may find that such a equivalence is consistent with the BR
 mappings in Eq. (\ref{eq:2Dmap-1a12})-(\ref{eq:2Dmap-4d12}).

Secondly, it is obvious that the left $mE_{\frac{1}{2}}^{1a}$ or $m^{\prime}E_{\frac{3}{2}}^{1a}$
WFs alone can not be gauged away because the BR generated from $mE_{\frac{1}{2}}^{1a}$ or $m^{\prime}E_{\frac{3}{2}}^{1a}$
can not be reproduced by any combination of WFs at other sites.

Thus, to complete the proof, we only need to prove a single rep
$E_{\frac{1}{2}}^{1a}+E_{\frac{3}{2}}^{1a}$ must
stay at $1a$ under any symmetric gauge transformation.
From the BR mappings, we find that $E_{\frac{1}{2}}^{1a}+E_{\frac{3}{2}}^{1a}$
can only be moved to $E_{\frac{1}{2}}^{1b}+E_{\frac{3}{2}}^{1b}$
or $E_{\frac{1}{2}}^{2c}$.
Such transformations are very unnatural from an intuitive perspective, because if
 we continuously move the four WFs at $1a$ to $1b$ or $2c$, the intermediate process
 will break either time-reversal or $C_{4}$.
(Enforced by the $C_{4}$ symmetry, the four WFs must move
 to four different directions, leading to separation of Kramers' pairs).
To prove this statement, here we show that the gauge transformation
 from $E_{\frac{1}{2}}^{1a}+E_{\frac{3}{2}}^{1a}$ to $E_{\frac{1}{2}}^{1b}+E_{\frac{3}{2}}^{1b}$
 or $E_{\frac{1}{2}}^{2c}$ must be singular.
Let us first consider the transformation from $E_{\frac{1}{2}}^{1a}+E_{\frac{3}{2}}^{1a}$ to $E_{\frac{1}{2}}^{1b}+E_{\frac{3}{2}}^{1b}$. For convenience,
 here we choose the WF bases with a ``cyclical'' gauge
\begin{equation}
\left|w_{1}\right\rangle =e^{-i\frac{\pi}{4}}\left|\frac{1}{2}\right\rangle +e^{-i\frac{\pi}{4}}\left|\bar{\frac{1}{2}}\right\rangle +e^{i\frac{\pi}{4}}\left|\frac{3}{2}\right\rangle +e^{i\frac{\pi}{4}}\left|\bar{\frac{3}{2}}\right\rangle \label{eq:w1}
\end{equation}
\begin{equation}
\left|w_{i+1}\right\rangle =C_{4}^{\prime}\left|w_{i}\right\rangle \qquad i=1,2,3\label{eq:wi}
\end{equation}
 where $C_{4}^{\prime}$ is the rotation operation centered at $1a$
 and $1b$ for the $E_{\frac{1}{2}}^{1a}+E_{\frac{3}{2}}^{1a}$ and $E_{\frac{1}{2}}^{1b}+E_{\frac{3}{2}}^{1b}$
 WFs, respectively.
In this guage, for both $1a$ and $1b$ positions,
$\left|w_{i}\right\rangle $ and $\left|w_{i+2}\right\rangle $ form
a Kramers' pair and the time-reversal rep matrix has the standard
form $\Omega$. According to Eq. (\ref{eq:Dk-def}) the $C_{4}$ sewing
matrices generated from $\left|w_{i}\right\rangle $ at $1a$ and
$1b$ positions can be derived respectively as
\begin{equation}
D^{\mathbf{k}}=W=\begin{bmatrix}0 & 0 & 0 & -1\\
1 & 0 & 0 & 0\\
0 & 1 & 0 & 0\\
0 & 0 & 1 & 0
\end{bmatrix}
\end{equation}
\begin{equation}
\tilde{D}^{\mathbf{k}}=We^{-ik_{y}}
\end{equation}
Now assume there is a well defined gauge transformation $U\left(\mathbf{k}\right)$
that relates these two sets of WFs, then from Eq. (\ref{eq:g-phi})
and (\ref{eq:T-phi}) such a gauge transform must satisfy
\begin{equation}
U\left(-\mathbf{k}\right)=\Omega U^{*}\left(\mathbf{k}\right)\Omega^{T}\label{eq:U-omg}
\end{equation}
\begin{equation}
U\left(C_{4}\mathbf{k}\right)=WU\left(\mathbf{k}\right)W^{\dagger}e^{ik_{y}}\label{eq:U-W}
\end{equation}
leading to the equation $U^{*}\left(\mathbf{k}\right)=U\left(\mathbf{k}\right)e^{ik_{x}+ik_{y}}$.
Thus we can write $U\left(\mathbf{k}\right)$ as an orthogonal
 matrix $O\left(\mathbf{k}\right)$ multiplied by a phase factor
\begin{equation}
U\left(\mathbf{k}\right)=O\left(\mathbf{k}\right)e^{-\frac{i}{2}\left(k_{x}+k_{y}\right)}\qquad O\left(\mathbf{k}\right)\in O\left(n\right)
\end{equation}
Substitute this back to Eq. (\ref{eq:U-omg}) and (\ref{eq:U-W}),
we get the constraints on $O\left(\mathbf{k}\right)$ as (i) $O\left(-\mathbf{k}\right)=\Omega O\left(\mathbf{k}\right)\Omega^{T}$
and (ii) $O\left(C_{4}\mathbf{k}\right)=WO\left(\mathbf{k}\right)W^{\dagger}$,
wherein the first constraint is implied by the second one. By requiring
$U\left(\mathbf{k}\right)$ to be periodic, we get another two constraints
as (iii) $O\left(k_{x}+2\pi,k_{y}\right)=-O\left(\mathbf{k}\right)$
and (iv) $O\left(k_{x},k_{y}+2\pi\right)=-O\left(\mathbf{k}\right)$.
In fact, such constraints make $O\left(\mathbf{k}\right)$ must be
singular at some momenta. To see this, express $O\left(\mathbf{k}\right)$
as
\begin{equation}
O\left(\mathbf{k}\right)=\xi\cdot\exp\left(-i\mathfrak{H}\left(\mathbf{k}\right)\right)
\end{equation}
where $\xi=\pm1$ is the determinant, and $\mathfrak{H}\left(\mathbf{k}\right)$
is a 4 by 4 imaginary Hermition matrix parameterized as
\begin{align}
\mathfrak{H}\left(\mathbf{k}\right) & =\omega\left(\mathbf{k}\right)\left[n_{1}\left(\mathbf{k}\right)\tau_{y}\sigma_{x}+n_{2}\left(\mathbf{k}\right)\tau_{0}\sigma_{y}+n_{3}\left(\mathbf{k}\right)\tau_{y}\sigma_{z}\right]\nonumber \\
 & +\theta\left(\mathbf{k}\right)\left[m_{1}\left(\mathbf{k}\right)\tau_{z}\sigma_{y}+m_{2}\left(\mathbf{k}\right)\tau_{x}\sigma_{y}+m_{3}\left(\mathbf{k}\right)\tau_{y}\sigma_{0}\right]
\end{align}
Here we take the convention that $\omega$, $\theta$ are real and positive,
 and $\mathbf{n}=\left(n_{1}n_{2}n_{3}\right)^{T}$, $\mathbf{m}=\left(m_{1}m_{2}m_{3}\right)^{T}$
 are unit vectors.
It should be noticed that as the first three matrices and the last
 three matrices respectively form a set of $SU\left(2\right)$ generators,
 and these two sets are commutative with each other, such a parameterization
 realizes an isomorphic mapping from $SO\left(4\right)$ to $SU\left(2\right)\times SU\left(2\right)$.
Therefore, the orthogonal matrix can be expressed as a product of
 two commutative matrices
\begin{equation}
O\left(\mathbf{k}\right)=\xi\cdot\mathcal{O}^{1}\left(\mathbf{k}\right)\mathcal{O}^{2}\left(\mathbf{k}\right)
\end{equation}
where
\begin{equation}
\mathcal{O}^{1}=\cos\omega-i\sin\omega\left(n_{1}\tau_{y}\sigma_{x}+n_{2}\tau_{0}\sigma_{y}+n_{3}\tau_{y}\sigma_{z}\right)
\end{equation}
\begin{equation}
\mathcal{O}^{2}=\cos\theta-i\sin\theta\left(m_{1}\tau_{y}\sigma_{x}+m_{2}\tau_{0}\sigma_{y}+m_{3}\tau_{y}\sigma_{z}\right)
\end{equation}
The  constraints (i) and (ii) lead to
\begin{equation}
\omega\left(C_{4}\mathbf{k}\right)=\omega\left(\mathbf{k}\right)\qquad\theta\left(C_{4}\mathbf{k}\right)=\theta\left(\mathbf{k}\right)\label{eq:omg-C4}
\end{equation}
\begin{equation}
\mathbf{n}\left(C_{4}\mathbf{k}\right)=\left[n_{2},n_{1},-n_{3}\right]^{T}\left(\mathbf{k}\right)
\end{equation}
\begin{equation}
\mathbf{m}\left(C_{4}\mathbf{k}\right)=\left[-m_{2},m_{1},m_{3}\right]^{T}\left(\mathbf{k}\right)\label{eq:nm-C4}
\end{equation}
And the anti-periodic property in constraints (iii) and (iv) must be realized
  by either $\mathcal{O}^{1}$ or $\mathcal{O}^{2}$.
In fact, whatever which $\mathcal{O}$ is chosen to be anti-periodic,
 the anti-periodic condition together with the symmetry constraints (Eq. (\ref{eq:omg-C4})-(\ref{eq:nm-C4})) will make $\mathcal{O}$ singular at some momenta.
Here we only take $\mathcal{O}^{1}$ as an example. With the anti-periodic condition,
only two branches of solutions exist, the first is
\begin{equation}
\omega\left(k_{x}+2\pi,k_{y}\right)=\omega\left(k_{x},k_{y}+2\pi\right)=\omega\left(\mathbf{k}\right)+\pi
\end{equation}
\begin{equation}
\mathbf{n}\left(k_{x}+2\pi,k_{y}\right)=\mathbf{n}\left(k_{x},k_{y}+2\pi\right)=\mathbf{n}\left(\mathbf{k}\right)
\end{equation}
and the second is
\begin{equation}
\omega\left(\mathbf{k}\right)=\frac{\pi}{2}
\end{equation}
\begin{equation}
\mathbf{n}\left(k_{x}+2\pi,k_{y}\right)=\mathbf{n}\left(k_{x},k_{y}+2\pi\right)=-\mathbf{n}\left(\mathbf{k}\right)\label{eq:n-anti}
\end{equation}
Obviously, the first breaks the $C_{4}$ symmetry constraint (Eq.
(\ref{eq:omg-C4})) because $\omega\left(\pi,0\right)=\omega\left(-\pi,0\right)+\pi$.
While, the second solution is singular at $\left(\pi0\right)$, since
there must be $\mathbf{n}\left(\pi,0\right)=0$ due to the $C_{4}$
constraint (Eq. (\ref{eq:omg-C4})) and the anti-periodic constraint
(Eq. (\ref{eq:n-anti})).
Therefore, we achieve the conclusion that there is no smooth gauge transformation can deform $E_{\frac{1}{2}}^{1a}+E_{\frac{3}{2}}^{1a}$
to $E_{\frac{1}{2}}^{1b}+E_{\frac{3}{2}}^{1b}$.
The proof for the inequivalence between $E_{\frac{1}{2}}^{1a}+E_{\frac{3}{2}}^{1a}$
 and $E_{\frac{1}{2}}^{2c}$ completely parallelizes the above process.

\subsection{Gauge invariant Wannier centers at other positions}
As $1a$ and $1b$ positions can be renamed to each other by re-choosing the origin
 point, there is no physical difference between them and all the statements about
 $1a$ should also hold for $1b$.
Therefore a set of WFs at $1b$ can be moved away by a symmetric gauge transformation
 \emph{only if} the WFs consist of an even number of rep $E_{\frac{1}{2}}^{1b}+E_{\frac{3}{2}}^{1b}$.
And, taking off all the even the number of  the rep $E_{\frac{1}{2}}^{1b}+E_{\frac{3}{2}}^{1b}$,
the centers of the left WFs are gauge invariant.

As for the $1c$ position, just like the $2E_{\frac{1}{2}}^{1a}+2E_{\frac{3}{2}}^{1a}$
 WFs in section \ref{sub:2DWF-1a}, a pair of $E_{\frac{1}{2}}^{2c}$
 can reduce to four Kramers' pairs at the $4d$ position, i.e. $2E_{\frac{1}{2}}^{2c}\sim E_{\frac{1}{2}}^{4d}$.
Thus, if there are an even number of $E_{\frac{1}{2}}^{2c}$ at the
$1c$ position all of them can be gauged away to $4d$ positions,
however, if there is an odd number of $E_{\frac{1}{2}}^{2c}$, at
least two WFs (a Kramers' pair) will stay at $1c$ under any symmetric
gauge transformation.

In summary, a set of WFs at any wyckoff position can be moved by symmetric gauge
 transformations \emph{if and only if} the rep (of the SSG at the Wyckoff position)
 they form is consistent with a set of $E_\frac{1}{2}^{4d}$ WFs.
In other words, all the move of WFs should pass through $4d$ positions, which is
very consistent with the intuitive picture.

\subsection{Occupied Wannier functions for the 2D model\label{sub:2DWF}}

The occupied bands of our 2D model give the BR
 $E_{\frac{1}{2}}^{\Gamma}+E_{\frac{3}{2}}^{\Gamma}+E_{\frac{1}{2}}^{M}+E_{\frac{3}{2}}^{M}$,
 which is irrelevant with the parameter $\Delta$ since the corresponding term vanish at $\Gamma$
 and $M$.
However, such a BR can not give a concrete real space information because it
 can be generated from $E_{\frac{1}{2}}^{1a}+E_{\frac{3}{2}}^{1a}$,
 or $E_{\frac{1}{2}}^{1b}+E_{\frac{3}{2}}^{1b}$, or $E_{\frac{1}{2}}^{2c}$.
Here, by constructing the WFs explicitly, we will show that the occupied
states are equivalent to $E_{\frac{1}{2}}^{1b}+E_{\frac{3}{2}}^{1b}$
WFs. We follow the projection procedure described in Ref. [\onlinecite{Soluyanov2011}].
Firstly, let us guess four trial local orbitals,
denoted by $\left|\gamma_{\alpha}\right\rangle $, at the home cell
and define them by the model orbitals
\begin{equation}
\left|\gamma_{\alpha}\right\rangle =\sum_{\mathbf{R}\beta}^{\prime}\left|a_{\beta\mathbf{R}}\right\rangle M_{\beta\alpha}^{\mathbf{R}}
\end{equation}
Here $\left|a_{\beta\mathbf{R}}\right\rangle $ is the $\beta$-th
atomic orbital in the lattice $\mathbf{R}$ in our 2D model, $M^{\mathbf{R}}$
(8 by 4) is the overlap matrices between atomic orbitals and the trial
orbitals, and the summation $\sum_{\mathbf{R}}^{\prime}$ is taken
only for a few lattices around the home cell. Assume $\left|\gamma_{\alpha}\right\rangle $
form the rep $E_{\frac{1}{2}}^{1b}+E_{\frac{3}{2}}^{1b}$  and limit the the summation over $\mathbf{R}$
within $\mathbf{R}_{1}=\left(00\right)$, $\mathbf{R}_{2}=\left(10\right)$,
$\mathbf{R}_{3}=\left(11\right)$, and $\mathbf{R}_{4}=(01)$,  the
symmetry properties satisfied by $\left|\gamma_{\alpha}\right\rangle $ imply a set of constraints on $M^{\mathbf{R}}$
\begin{equation}
M^{\mathbf{R}_{i+1}}=C_{4}M^{\mathbf{R}_{i}}D^{\gamma\dagger}\left(C_{4}\right)\label{eq:MR-C4}
\end{equation}
\begin{equation}
M^{\mathbf{R}_{i}}=TM^{\mathbf{R}_{i}}\Omega^{T}\label{eq:MR-T}
\end{equation}
Here $C_{4}=\tau_{z}e^{-i\frac{\pi}{4}s_{z}}$ is the
$C_{4}$ operator on the atomic orbitals, $D^{\gamma}\left(C_{4}\right)$
is the $E_{\frac{1}{2}}^{1b}+E_{\frac{3}{2}}^{1b}$ rep
matrix of $C_{4}$, $T=-is_{y}K$
is the time-reversal operator on the atomic orbitals, and $\Omega$ is
the time-reversal rep on $E_{\frac{1}{2}}^{1b}+E_{\frac{3}{2}}^{1b}$.
Aligning the gauge of occupied Bloch states with respect to the trial orbitals,
we can define a set of projected Bloch-like states
\begin{equation}
\left|\Upsilon_{\alpha\mathbf{k}}\right\rangle =\sum_{n\in\mathrm{occ}}\left|\psi_{n\mathbf{k}}\right\rangle \left\langle \psi_{n\mathbf{k}}\right|\gamma_{\alpha}\rangle
\end{equation}
and the overlap matrix between them
\begin{equation}
S_{\alpha\beta}\left(\mathbf{k}\right)=\langle\Upsilon_{\alpha\mathbf{k}}|\Upsilon_{\beta\mathbf{k}}\rangle
\end{equation}
Then by the L$\ddot{\text{o}}$wdin orthonormalization procudure,
a set of orthonormal Bloch-like states are obtained
\begin{equation}
\left|\tilde{\psi}_{\alpha\mathbf{k}}\right\rangle =\sum_{\beta}S_{\beta\alpha}^{-\frac{1}{2}}\left(\mathbf{k}\right)\left|\Upsilon_{\beta\mathbf{k}}\right\rangle
\end{equation}
The WFs transformed from these Bloch-like states will
be well localized and satisfy the $E_{\frac{1}{2}}^{1b}+E_{\frac{3}{2}}^{1b}$
rep as long as the overlap matrix $S\left(\mathbf{k}\right)$
is non-singular over the whole Brillouin zone.

In practice, we generate a random $M^{\mathbf{R}_{1}}$ matrix and
symmetrize it due to Eq. (\ref{eq:MR-C4})-(\ref{eq:MR-T}).
A non-singular $S\left(\mathbf{k}\right)$ has been successfully obtained.
The WCs are also confirmed by the wilson loop method and are found indeed to locate at the $1b$ position.
In the wilson loop method, the center of the $\alpha$-th WF is calculated
by
\begin{equation}
x_{\alpha}=\frac{1}{2\pi}\int dk_{y}x_{\alpha}\left(k_{y}\right)
\end{equation}
\begin{equation}
y_{\alpha}=\frac{1}{2\pi}\int dk_{x}y_{\alpha}\left(k_{x}\right)
\end{equation}
\begin{align}
e^{i2\pi x_{\alpha}\left(k_{y}\right)} & =\langle\tilde{u}_{\alpha,0,k_{y}}|\tilde{u}_{\alpha,\left(N-1\right)\Delta k,k_{y}}\rangle\cdots\nonumber \\
 & \times\langle\tilde{u}_{\alpha,2\Delta k,k_{y}}|\tilde{u}_{\alpha,\Delta k,k_{y}}\rangle\langle\tilde{u}_{\alpha,\Delta k,k_{y}}|\tilde{u}_{\alpha,0,k_{y}}\rangle
\end{align}
\begin{align}
e^{i2\pi y_{\alpha}\left(k_{x}\right)} & =\langle\tilde{u}_{\alpha,k_{y},0}|\tilde{u}_{\alpha,k_{x},\left(N-1\right)\Delta k}\rangle\cdots\nonumber \\
 & \times\langle\tilde{u}_{\alpha,k_{x},2\Delta k}|\tilde{u}_{\alpha,k_{x},\Delta k}\rangle\langle\tilde{u}_{\alpha,k_{x},\Delta k}|\tilde{u}_{\alpha,k_{x},0}\rangle
\end{align}
where $\Delta k=\frac{2\pi}{N}$, and $\left|\tilde{u}_{\alpha\mathbf{k}}\right\rangle $
is the periodic part of the orthonormal Bloch-like state $\left|\tilde{\psi}_{\alpha\mathbf{k}}\right\rangle $.

It should be noticed that, whatever the value of $\Delta$ takes, the occupied WCs should locate at
 $1b$.
This is simply because the $\Delta$ term can not close the band gap and the four WFs $E_\frac{1}{2}^{1b}+E_\frac{3}{2}^{1b}$ can not be
 moved away from $1b$ by any adiabatic process, as proved before.

\section{Wannier center flow in 3D system}

\subsection{Classification of the flows}\label{sub:flowclass}

For a 3D system with both time-reversal and $C_{4}$ symmetries, we can choose
a tetragonal cell with its principal axis along the $z$ direction
and apply a fourier transformation along $z$.
Here we focus on the case where all the $k_{z}$-slices are equivalent to some 2D atomic insulator.
The above conclusions about the gauge invariant 2D WCs applies for the $k_{z}=0,\pi$-slices because both the time-reversal and $C_{4}$ symmetries present there.
However, in general intermediate $k_{z}$-slices, the above conclusions fail because of the absence of time-reversal symmetry.
Then an immediate observation follows is that a bulk state must be  nontrivial if there is a mismatch between its 2D WCs in $k_{z}=0$- and $k_{z}=\pi$-slices.
We argue that the bulk topology is only determined by the WFs at the $k_{z}=0,\pi$-slices, because, as the sewing matrices in the intermediate slices are completely determined by the two ends (from the compatibility
relation) and all the 2D atomic insulator with same sewing matrices are topologically equivalent, all the possible evolutions must be equivalent with each other.
Therefore, for a given set of 2D WFs at the $k_{z}=0$- and $k_{z}=\pi$-slices, a nontrivial flow exists if (i) the WCs at $k_{z}=0$- and $k_{z}=\pi$-slices are inequivalent with each other and (ii) the WCs at the two ends can be \emph{continuously} deformed to each other by a time-reversal breaking process.
Considering that the gauge invariant WCs at $k_{z}=0,\pi$-slices
can only locate at $1a$, $1b$ and $2c$ positions and the
flows between $1a$ and $2c$ can be generated form the flows between
$1b$ and $2c$ and the flows between $1b$ and $1a$, to figure out
the flow classification we only need to discuss the latter two cases.

Let us start with the flows between $1b$ and $1a$. For both $1b$
and $1a$ in $k_{z}=0$- or $k_{z}=\pi$-slices, we only need to study
the left immovable irreps
\begin{equation}
n\left(E_{\frac{1}{2}}+E_{\frac{3}{2}}\right)+mE_{\frac{1}{2}}+m^{\prime}E_{\frac{3}{2}}
\end{equation}
as discussed in section \ref{sub:2DWF-1a}. Here
$n=0,1$ and the one of $m,m^{\prime}$ equals to zero. Firstly, we
will show that the $2m$ ($2m^{\prime}$) 2D WFs in the $E_{\frac{1}{2}}$
($E_{\frac{3}{2}}$) irreps at $1a$ or $1b$
can not move along the flow. This can be seen by presuming an
infinite small move and comparing the $C_{4}$ rep  matrix
before and after the move. Here we take $m$ $E_{\frac{1}{2}}^{1a}$
irreps for an example. Before the move, the $C_{4}$ rep
matrix is a direct sum of $m$ $E_{\frac{1}{2}}$ rep matrices,
thus the trace gives $\mathrm{Tr}\left[D\left(C_{4}\right)\right]=m\sqrt{2}$.
While, after the move, the WFs locating at $4d$ positions must form
a traceless $D\left(C_{4}\right)$, because the four equivalent $4d$
positions transform to each other in turn under the $C_{4}$ rotation.
Therefore the presumption of the infinite small move is untenable.
Secondly, notice that a single rep $E_{\frac{1}{2}}+E_{\frac{3}{2}}$
can be separated into four WFs at $4d$ positions by a time-reversal breaking process, which can be achieved in two steps. In the first step, we
take the ``cyclical'' gauge defined in Eq. (\ref{eq:w1})-(\ref{eq:wi}),
where $C_{4}$ transforms the WFs to each other in turn and time-reversal transforms
$\left|w_{i}\right\rangle $ to $\left|w_{i+2}\right\rangle $. In
the second step, we split $\left|w_{1}\right\rangle $ and $\left|w_{3}\right\rangle $
in the $x$ direction and split the $\left|w_{2}\right\rangle $ and
$\left|w_{4}\right\rangle $ in the $y$ direction.
Thus, through the
$4d$ positions, there can be a flow $E_{\frac{1}{2}}^{1b}+E_{\frac{3}{2}}^{1b}\to E_{\frac{1}{2}}^{1a}+E_{\frac{3}{2}}^{1a}$
or $E_{\frac{1}{2}}^{1a}+E_{\frac{3}{2}}^{1a}\to E_{\frac{1}{2}}^{1b}+E_{\frac{3}{2}}^{1b}$,
where the arrow represents a flow from $k_{z}=0$ to $k_{z}=\pi$.
Such flows are gauge invariant because both $E_{\frac{1}{2}}^{1b}+E_{\frac{3}{2}}^{1b}$
and $E_{\frac{1}{2}}^{1a}+E_{\frac{3}{2}}^{1a}$ at the two ends can
not be gauged away.
However, double of the flows must be trivial because the $2E_{\frac{1}{2}}^{1a}+2E_{\frac{3}{2}}^{1a}$
or the $2E_{\frac{1}{2}}^{1b}+2E_{\frac{3}{2}}^{1b}$ can be gauged away,
as discussed in section \ref{sub:2DWF-1a}. It should also be noticed
that the flow $E_{\frac{1}{2}}^{1a}+E_{\frac{3}{2}}^{1a}\to E_{\frac{1}{2}}^{1b}+E_{\frac{3}{2}}^{1b}$
is equal to the flow $E_{\frac{1}{2}}^{1b}+E_{\frac{3}{2}}^{1b}\to E_{\frac{1}{2}}^{1a}+E_{\frac{3}{2}}^{1a}$
module a trivial flow $2E_{\frac{1}{2}}^{1a}+2E_{\frac{3}{2}}^{1a}\to2E_{\frac{1}{2}}^{1b}+2E_{\frac{3}{2}}^{1b}$.
Therefore, in summary, we obtain a $\mathbb{Z}_{2}$ class of the
flow between $1b$ and $1a$, wherein the nontrivial element
$E_{\frac{1}{2}}^{1b}+E_{\frac{3}{2}}^{1b}\to E_{\frac{1}{2}}^{1a}+E_{\frac{3}{2}}^{1a}$
manifests itself by 1D helical modes, as discussed in the text.

The discussion of the flow between $1b$ and $2c$ is much more simple.
As discussed above, at $1b$ position, a nontrivial flow must
start or end with $E_{\frac{1}{2}}^{1b}+E_{\frac{3}{2}}^{1b}$ irreps.
While, at the side of $2c$, the flow can only merge to the $E_{\frac{1}{2}}^{2c}$
irrep. As the
$C_{4}$ sewing matrix of $E_{\frac{1}{2}}^{2c}$ irrep is consistent
with the $C_{4}$ sewing matrix of $E_\frac{1}{2}^{4d}$, the
flow $E_{\frac{1}{2}}^{1b}+E_{\frac{3}{2}}^{1b}\to E_{\frac{1}{2}}^{2c}$
indeed can be realized by a time-reversal breaking process. Similar with the
$1b\to1a$ flow, this flow also generates a $\mathbb{Z}_{2}$ class,
because the double of it $2E_{\frac{1}{2}}^{1b}+2E_{\frac{3}{2}}^{1b}\to2E_{\frac{1}{2}}^{2c}$
can be trivialized by deformations at
the two ends.
What kind of surface state is manifested in this class of flow?
By an isomorphic mapping from the flow to the surface dispersion,
we find even number Dirac points on both the $zx$ and the $yz$ surfaces,
indicating the bulk is a weak topological insulator.

Beware that, the two flows defined above do not give a complete classification
of the time-reversal and $C_{4}$ protected 3D topological crystalline insulators,
instead, they only classify the a special kind of insulators where
each $k_{z}$-slice is equivalent with a 2D atomic insulator.

\subsection{Construct the flow of the 3D model}\label{sub:ModelFlow}

\begin{figure}
\begin{centering}
\includegraphics[width=0.7\linewidth]{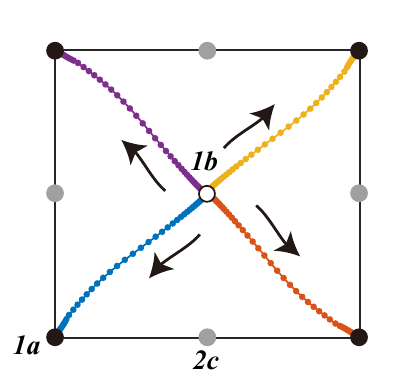}
\par\end{centering}
\protect\caption{\label{fig:Flow} The numerically calculated WC flow from $k_z=0$- to $k_z=\pi$-slices in the 3D model,
 where the parameter $\Delta$ is set to 0.2.}
\end{figure}

To verify our theory, in this section we explicitly show the $1b\to1a$
flow in our 3D model by constructing 2D WFs continuously from $k_{z}=0$-
to $k_{z}=\pi$-slices. As described in section \ref{sub:2DWF}, the
overlap matrix in real space ($M^{\mathbf{R}}$) can be thought as the
input of the construction procedure. Therefore, to get continuous WF
gauges from $k_{z}=0$ to $k_{z}=\pi$, we can firstly work out the
overlap matrices at the two ends, i.e. $M^{\mathbf{R}}\left(0\right)$
and $M^{\mathbf{R}}\left(\pi\right)$, and then interpolate $M^{\mathbf{R}}\left(k_{z}\right)$
in the intermediate slices. As the $k_{z}=0$-slice is equivalent
with our 2D model, we can directly use the 2D WFs constructed in section
\ref{sub:2DWF}. While, to be consistent with the flow process, here
we choose the ``cyclical'' gauge defined in Eq. (\ref{eq:w1})-(\ref{eq:wi}), for which the overlap matrices build for $E_{\frac{1}{2}}^{1b}+E_{\frac{3}{2}}^{1b}$
in appendix \ref{sub:2DWF}, denoted as $\tilde{M}^{\mathbf{R}}$ here,
should be multiplied by an unitary matrix
\begin{equation}
M^{\mathbf{R}_{i}}\left(0\right)=\tilde{M}^{\mathbf{R}_{i}}V^{\dagger}
\end{equation}
where $V$ can be read from the ``cyclical'' gauge definition
in Eq. (\ref{eq:w1})-(\ref{eq:wi}). The overlap matrix at $k_{z}=\pi$
can also be constructed in a similar way: randomly generate an overlap matrix
and then symmetrize it. To be consistent with the flow, in $k_{z}=\pi$-slice
we also choose the ``cyclical" gauge for the four $E_\frac{1}{2}^{1a} + E_\frac{3}{2}^{1a}$
trial orbitals and put them in the lattices $\mathbf{R}^{1}=\left(00\right)$,
$\mathbf{R}^{2}=\left(10\right)$, $\mathbf{R}^{3}=\left(11\right)$,
and $\mathbf{R}^{4}=\left(01\right)$, respectively. We also generate
an additional $\delta M^{\mathbf{R}_{i}}$ term to cover the time-reversal
breaking process in the intermediate slices. At last, the overlap matrix can
be interpolated as
\begin{align}
M^{\mathbf{R}_{i}}\left(k_{z}\right) & =\left(1-\frac{k_{z}}{\pi}\right)M^{\mathbf{R}_{i}}\left(0\right)+\frac{k_{z}}{\pi}M^{\mathbf{R}_{i}}\left(\pi\right)\nonumber \\
 & +\lambda \frac{k_z}{\pi}\left(1-\frac{k_{z}}{\pi}\right)\delta M^{\mathbf{R}_{i}}
\end{align}
Here $\lambda$ is an adjustable parameter. Within the continuous
symmetric WF gauges from $k_{z}=0$ to $k_{z}=\pi$, we calculate
the evolution of WCs by the Wilson loop method and plotted
it in Fig. (\ref{fig:Flow}), which indeed coincides with the nontrivial $\mathbb{Z}_2$-flow.

\section{Low energy theory}\label{sub:kp}

Another perspective to understand the 1D helical modes is from the
effective low energy theory on the surfaces. As the 3D model can
be thought as two copies of topological insulators plus a mixing term, on
surfaces there should be two Dirac points and a mass term between
them
\begin{equation}
H=k_{1}\tau_{0}\tilde{s}_{1}+k_{2}\tau_{0}\tilde{s}_{2}+m\tau_{y}\tilde{s}_{3}\label{eq:Hsurf}
\end{equation}
Here $k_{1}$, $k_{2}$ are the surface momenta, $\tilde{s}_{i}$
are pauli matrices representing the pseudo spin. In the geometry in Fig.
(3) in the text, we set $k_{1}=k_{y}$, $k_{2}=k_{z}$ for the
$yz$ surface, and $k_{1}=-k_{x}$, $k_{2}=k_{z}$ for the $zx$ surface.
For a general surface deviating from the $yz$ plane by an angle $\theta$,
we set
\begin{equation}
k_{1}=-\sin\theta k_{x}+\cos\theta k_{y}
\end{equation}
\begin{equation}
k_{2}=k_{z}
\end{equation}
Then, by the symmetry analysis in the following, we show that the
mass term flips its sign under the $C_{4}$ rotation, i.e. $m\left(\theta\right)=-m\left(\theta+\frac{\pi}{2}\right)$.
Thus, enforced by the $C_{4}$ symmetry, there must be domain walls
of masses, where the helical states live, on the surfaces. Two interesting
observations immediately follow. The first is that, as the domain
walls are enforced by the $C_{4}$ symmetry, in general, they do not
necessarily locate at the hinges. Instead, it can be anywhere
on the surfaces. For our 3D model, the helical modes are pinned at the hinges by the accidental mirror symmetry.
The second observation is that, the 1D helical modes are stable against to any time-reversal preserving
perturbations---even the $C_{4}$ breaking perturbations---because
to gap out the helical modes one need to move two domain walls separated far away in real space together to annihilate them in pair, which can not be realized by a perturbation.

Now let us derive the effective theory in Eq. (\ref{eq:Hsurf}) and
prove the mass flipping under $C_{4}$. Firstly, we write the bulk
Hamiltonian in the surface coordinates $\left(k_{1}k_{2}k_{3}\right)$,
where $k_{3}=\cos\theta k_{x}+\sin\theta k_{y}$, and expand it to
first order of $k_{1}$, $k_{2}$ and second order of $k_{3}$
\begin{align}
H & = \left(-1+\frac{k_{3}^{2}}{2}\right) \tau_{0}\sigma_{z}s_{0} + k_{3}\tau_{0}\sigma_{x}\left(s_{x}\cos\theta+s_{y}\sin\theta\right)\nonumber \\
 & +k_{1}\tau_{0}\sigma_{x}\left(-s_{x}\sin\theta+s_{y}\cos\theta\right)+k_{2}\tau_{0}\sigma_{x}s_{z}\nonumber \\
 & +\frac{\Delta\left(\theta\right)}{2}k_{3}^{2}\tau_{y}\sigma_{y}s_{0}
\end{align}
where
\begin{equation}
\Delta\left(\theta\right)=\Delta\left(\sin^{2}\theta-\cos^{2}\theta\right)
\end{equation}
It should be noticed that, even this is a low energy theory, the property
$\Delta\left(\theta+\frac{\pi}{2}\right)=-\Delta\left(\theta\right)$
holds to any order because this is enforced by the symmetry relation
$C_{4}\tau_{y}\tau_{y}s_{0}C_{4}^{-1}=-\tau_{y}\tau_{y}s_{0}$. By
the gauge transformation $s_{x}\cos\theta+s_{y}\sin\theta\to s_{3}$,
$-s_{x}\sin\theta+s_{y}\cos\theta\to s_{1}$, $s_{z}\to s_{2}$,
and the replacement $k_{3}\to-i\partial_{3}$, we get the
Hamiltonian on the surface as
\begin{align}
H & =\left(-1-\frac{\partial_{3}^{2}}{2}\right)\tau_{0}\sigma_{z}s_{0}-i\partial_{3}\tau_{0}\sigma_{x}s_{3}-\frac{\Delta\left(\theta\right)}{2}\partial_{3}^{2}\tau_{y}\sigma_{y}s_{0}\nonumber \\
 & +k_{1}\tau_{0}\sigma_{x}s_{1}+k_{2}\tau_{0}\sigma_{x}s_{2}\label{eq:Heff-bulk}
\end{align}
Then, neglecting the $k_{1}$, $k_{2}$, and $\Delta$ terms, we get
the zero modes equation in the $x_{3}\ge 0$ semi-infinite system
\begin{equation}
\left[\left(1+\frac{1}{2}\partial_{x}^{2}\right)-\tau_{0}\sigma_{y}s_{3}\partial_{3}\right]\psi\left(x_{3}\right)=0
\end{equation}
with the boundary condition
\begin{equation}
\psi\left(0\right)=\psi\left(\infty\right)=0
\end{equation}
It has four solutions
\begin{equation}
\psi_{\mu n}\left(x_{3}\right)=u_{\mu n}\frac{1}{\sqrt{\mathcal{C}}}\left(e^{-\lambda_{+}x_{3}}-e^{-\lambda_{-}x_{3}}\right)
\end{equation}
, where
\begin{equation}
\lambda_{\pm}=1\pm i
\end{equation}
\begin{equation}
u=\frac{1}{\sqrt{2}}\begin{bmatrix}i & 0 & 0 & 0\\
0 & i & 0 & 0\\
1 & 0 & 0 & 0\\
0 & -1 & 0 & 0\\
0 & 0 & i & 0\\
0 & 0 & 0 & i\\
0 & 0 & 1 & 0\\
0 & 0 & 0 & -1
\end{bmatrix}
\end{equation}
Then expand the remaining terms in Eq. (\ref{eq:Heff-bulk}) on these
solutions, we get the effective theory
\begin{align}
\mathcal{H} & =k_{1}\tau_{0}\tilde{s}_{1}+k_{2}\tau_{0}\tilde{s}_{2}+m\left(\theta\right)\tau_{y}\tilde{s}_{3}
\end{align}
where $m\left(\theta\right)$ is the mass induced by the mixing term
$\Delta\left(\theta\right)$
\begin{align}
m\left(\theta\right) & =\frac{\Delta\left(\theta\right)}{2\mathcal{C}}\int dx_{3}\left(e^{-\lambda_{+}x_{3}}-e^{-\lambda_{-}x_{3}}\right)^{*}\nonumber \\
 & \qquad\times\partial_{3}^{2}\left(e^{-\lambda_{+}x_{3}}-e^{-\lambda_{-}x_{3}}\right)
\end{align}
Therefore, as $\Delta\left(\theta+\frac{\pi}{2}\right)=-\Delta\left(\theta\right)$,
the sign of mass must flip under the $C_{4}$ rotation.

\section{Symmetry indicators} \label{sub:indicator}

The $\mathbb{Z}_2$-flow has provided a good physical picture and serves
as a topological invariant characterizing the nontrivial states. However,
for real materials, it is practically impossible to find smooth and symmetric
gauges to calculate the flow. Thus it would be very useful if there
is some Fu-Kane-like criterion that can diagnose the topology from merely symmetry eigenvalues.
As will be shown
latter, such a criterion indeed exists if the system has an additional
inversion symmetry.

We follow the newly developed symmetry indicator method \cite{Po2017,Bradlyn2017} to diagnose
the topology.
In this method, a BR is represented by a column vector of integers,
 where each entry gives the appeared number of a particular  irrep at a particular high-symmetry momentum.
All the admissible BRs that satisfy the compatibility relations form a linear space.
On the other hand, the bases of this linear space
 can also be generated from a set of atomic insulators.
Consequently, any BR can be expanded by the atomic BR bases with integral or fractional coefficients, wherein fractional coefficients imply some kind of nontrivial topology.

\subsection{Symmetry indicators in space group $P4/m$}

\begin{table*}
\begin{tabular}{|c|c|c|c|c|}
\hline
\multirow{2}{*}{Site sym.} & \multicolumn{2}{c|}{Wyckoff position} & \multicolumn{2}{c|}{High sym. K}\tabularnewline
\cline{2-5}
 & W & Irrep & K & Irrep\tabularnewline
\hline
\hline
\multirow{4}{*}{$C_{4h}$} & $1a$ $\left(000\right)$ & \multirow{4}{*}{$E_{\frac{3}{2}u}$, $E_{\frac{3}{2}g}$, $E_{\frac{1}{2}u}$, $E_{\frac{1}{2}g}$} & $\Gamma$ $\left(000\right)$ & \multirow{4}{*}{$E_{\frac{3}{2}u}$, $E_{\frac{3}{2}g}$, $E_{\frac{1}{2}u}$, $E_{\frac{1}{2}g}$}\tabularnewline
\cline{2-2} \cline{4-4}
 & $1b$ $\left(00\frac{1}{2}\right)$ &  & $Z$ $\left(00\pi\right)$ & \tabularnewline
\cline{2-2} \cline{4-4}
 & $1c$ $\left(\frac{1}{2}\frac{1}{2}0\right)$ &  & $M$ $\left(\pi\pi0\right)$ & \tabularnewline
\cline{2-2} \cline{4-4}
 & $1d$ $\left(\frac{1}{2}\frac{1}{2}\frac{1}{2}\right)$ &  & $A$ $\left(\pi\pi\pi\right)$ & \tabularnewline
\hline
\multirow{2}{*}{$C_{2h}$} & $2e$ $\left(\frac{1}{2}00\right)$ & \multirow{2}{*}{$E_{\frac{1}{2}u}$, $E_{\frac{1}{2}g}$} & $X$ $\left(0\pi0\right)$ & \multirow{2}{*}{$E_{\frac{1}{2}u}$, $E_{\frac{1}{2}g}$}\tabularnewline
\cline{2-2} \cline{4-4}
 & $2f$ $\left(\frac{1}{2}0\frac{1}{2}\right)$ &  & $R$ $\left(0\pi\pi\right)$ & \tabularnewline
\hline
\multirow{2}{*}{$C_{4}$} & $2g$ $\left(00z\right)$ & \multirow{2}{*}{$E_{\frac{3}{2}}$, $E_{\frac{1}{2}}$} & $\Lambda$ $\left(00u\right)$ & \multirow{2}{*}{$E_{\frac{3}{2}}$, $E_{\frac{1}{2}}$}\tabularnewline
\cline{2-2} \cline{4-4}
 & $2h$ $\left(\frac{1}{2}\frac{1}{2}z\right)$ &  & $V$ $\left(\pi\pi u\right)$ & \tabularnewline
\hline
$C_{2}$ & $4i$ $\left(\frac{1}{2}0z\right)$ & $E_{\frac{1}{2}}$ & $W$ $\left(0\pi u\right)$ & $E_{\frac{1}{2}}$\tabularnewline
\hline
\multirow{2}{*}{$C_{s}$} & $4j$ $\left(xy0\right)$ & \multirow{2}{*}{$E_{\frac{1}{2}}$} & $D$ $\left(uv0\right)$ & \multirow{2}{*}{$E_{\frac{1}{2}}$}\tabularnewline
\cline{2-2} \cline{4-4}
 & $4k$ $\left(xy\frac{1}{2}\right)$ &  & $E$ $\left(uv\pi\right)$ & \tabularnewline
\hline
$C_{1}$ & $8I$ $\left(xyz\right)$ & $E_{\frac{1}{2}}$ & $GP$ & $E_{\frac{1}{2}}$\tabularnewline
\hline
\end{tabular}

\protect\caption{\label{tab:P4/m-wkf} Wyckoff positions, high-symmetry momenta, and
 the irreps of their SSGs in space group $P4/m$ (with time-reversal symmetry).}
\end{table*}

\begin{table*}
\begin{tabular}{|c|c|c|}
\hline
 &  & BR \tabularnewline
\hline
\hline
$\mathbf{A}_{1}$ & $E_{\frac{1}{2}g}^{1a}$ & $E_{\frac{1}{2}g}^{\Gamma}+E_{\frac{1}{2}g}^{M}+E_{\frac{1}{2}g}^{X}+E_{\frac{1}{2}g}^{Z}+E_{\frac{1}{2}g}^{A}+E_{\frac{1}{2}g}^{R}$\tabularnewline
\hline
$\mathbf{A}_{2}$ & $E_{\frac{3}{2}g}^{1a}$ & $E_{\frac{3}{2}g}^{\Gamma}+E_{\frac{3}{2}g}^{M}+E_{\frac{1}{2}g}^{X}+E_{\frac{3}{2}g}^{Z}+E_{\frac{3}{2}g}^{A}+E_{\frac{1}{2}g}^{R}$\tabularnewline
\hline
$\mathbf{A}_{3}$ & $E_{\frac{1}{2}u}^{1a}$ & $E_{\frac{1}{2}u}^{\Gamma}+E_{\frac{1}{2}u}^{M}+E_{\frac{1}{2}u}^{X}+E_{\frac{1}{2}u}^{Z}+E_{\frac{1}{2}u}^{A}+E_{\frac{1}{2}u}^{R}$\tabularnewline
\hline
$\mathbf{A}_{4}$ & $E_{\frac{3}{2}u}^{1a}$ & $E_{\frac{3}{2}u}^{\Gamma}+E_{\frac{3}{2}u}^{M}+E_{\frac{1}{2}u}^{X}+E_{\frac{3}{2}u}^{Z}+E_{\frac{3}{2}u}^{A}+E_{\frac{1}{2}u}^{R}$\tabularnewline
\hline
$\mathbf{A}_{5}$ & $E_{\frac{1}{2}g}^{1b}$ & $E_{\frac{1}{2}g}^{\Gamma}+E_{\frac{1}{2}g}^{M}+E_{\frac{1}{2}g}^{X}+E_{\frac{1}{2}u}^{Z}+E_{\frac{1}{2}u}^{A}+E_{\frac{1}{2}u}^{R}$\tabularnewline
\hline
$\mathbf{A}_{6}$ & $E_{\frac{3}{2}g}^{1b}$ & $E_{\frac{3}{2}g}^{\Gamma}+E_{\frac{3}{2}g}^{M}+E_{\frac{1}{2}g}^{X}+E_{\frac{3}{2}u}^{Z}+E_{\frac{3}{2}u}^{A}+E_{\frac{1}{2}u}^{R}$\tabularnewline
\hline
$\mathbf{A}_{7}$ & $E_{\frac{1}{2}g}^{1c}$ & $E_{\frac{1}{2}g}^{\Gamma}+E_{\frac{3}{2}g}^{M}+E_{\frac{1}{2}u}^{X}+E_{\frac{1}{2}g}^{Z}+E_{\frac{3}{2}g}^{A}+E_{\frac{1}{2}u}^{R}$\tabularnewline
\hline
$\mathbf{A}_{8}$ & $E_{\frac{3}{2}g}^{1c}$ & $E_{\frac{3}{2}g}^{\Gamma}+E_{\frac{1}{2}g}^{M}+E_{\frac{1}{2}u}^{X}+E_{\frac{3}{2}g}^{Z}+E_{\frac{1}{2}g}^{A}+E_{\frac{1}{2}u}^{R}$\tabularnewline
\hline
$\mathbf{A}_{9}$ & $E_{\frac{1}{2}u}^{1c}$ & $E_{\frac{1}{2}u}^{\Gamma}+E_{\frac{3}{2}u}^{M}+E_{\frac{1}{2}g}^{X}+E_{\frac{1}{2}u}^{Z}+E_{\frac{3}{2}u}^{A}+E_{\frac{1}{2}g}^{R}$\tabularnewline
\hline
$\mathbf{A}_{10}$ & $E_{\frac{1}{2}g}^{1d}$ & $E_{\frac{1}{2}g}^{\Gamma}+E_{\frac{3}{2}g}^{M}+E_{\frac{1}{2}u}^{X}+E_{\frac{1}{2}u}^{Z}+E_{\frac{3}{2}u}^{A}+E_{\frac{1}{2}g}^{R}$\tabularnewline
\hline
$\mathbf{A}_{11}$ & $E_{\frac{3}{2}g}^{1d}$ & $E_{\frac{3}{2}g}^{\Gamma}+E_{\frac{1}{2}g}^{M}+E_{\frac{1}{2}u}^{X}+E_{\frac{3}{2}u}^{Z}+E_{\frac{1}{2}u}^{A}+E_{\frac{1}{2}g}^{R}$\tabularnewline
\hline
$\mathbf{A}_{12}$ & $E_{\frac{1}{2}g}^{2e}$ & $E_{\frac{1}{2}g}^{\Gamma}+E_{\frac{3}{2}g}^{\Gamma}+E_{\frac{1}{2}u}^{M}+E_{\frac{3}{2}u}^{M}+E_{\frac{1}{2}g}^{X}+E_{\frac{1}{2}u}^{X}+E_{\frac{1}{2}g}^{Z}+E_{\frac{3}{2}g}^{Z}+E_{\frac{1}{2}u}^{A}+E_{\frac{3}{2}u}^{A}+E_{\frac{1}{2}g}^{R}+E_{\frac{1}{2}u}^{R}$\tabularnewline
\hline
$\mathbf{A}_{13}$ & $E_{\frac{1}{2}g}^{2f}$ & $E_{\frac{1}{2}g}^{\Gamma}+E_{\frac{3}{2}g}^{\Gamma}+E_{\frac{1}{2}u}^{M}+E_{\frac{3}{2}u}^{M}+E_{\frac{1}{2}g}^{X}+E_{\frac{1}{2}u}^{X}+E_{\frac{1}{2}u}^{Z}+E_{\frac{3}{2}u}^{Z}+E_{\frac{1}{2}g}^{A}+E_{\frac{3}{2}g}^{A}+E_{\frac{1}{2}g}^{R}+E_{\frac{1}{2}u}^{R}$\tabularnewline
\hline
\end{tabular}

\protect\caption{\label{tab:P4/m-AI} Atomic BR bases of space group $P4/m$. In the first, second, and third columns,
 we list the notations of atomic BR bases, the irreps in real space to generate it,
 and the BR in momentum space, respectively.}

\end{table*}

The smallest space group containing both $C_{4}$ and inversion is $P4/m$,
 whose symmetry indicators form a group $\mathbb{Z}_{2}\times\mathbb{Z}_{4}\times\mathbb{Z}_{8}$.
In this section, we will work out the generators of the group.

In table \ref{tab:P4/m-wkf}, we list all the wyckoff positions and high-symmetry momenta,
 and the irreps of their SSGs in space group $P4/m$.
By definition, a BR should be given by the numbers of irreps at $\Gamma$, $M$, $X$,
 $Z$, $A$, $R$, $\Lambda$, $V$, $W$, $D$, $E$.
However, as the latter five momenta can be continuously connected to the former six momenta
 which have higher symmetries, the irreps at these five momenta can be inferred directly from the knowledge
 of the irreps at the former six momenta and the compatibility relation.
Thus we conclude that the BR is completely determined by the irreps at $\Gamma$, $M$, $X$,
 $Z$, $A$, and $R$.
By applying Eq. (\ref{eq:Dk-def}) repeatedly, we have found all the independent atomic BR bases and summarize
 them in table \ref{tab:P4/m-AI}.

Now let us find out all the the compatibility relations allowed BRs.
The compatibility relations consist of  five constraints on the occupation number
\begin{eqnarray}		&  & n\left(E_{\frac{1}{2}g}^{\Gamma}\right)+n\left(E_{\frac{3}{2}g}^{\Gamma}\right)+n\left(E_{\frac{1}{2}u}^{\Gamma}\right)+n\left(E_{\frac{3}{2}u}^{\Gamma}\right)\nonumber \\
 & = & n\left(E_{\frac{1}{2}g}^{M}\right)+n\left(E_{\frac{3}{2}g}^{M}\right)+n\left(E_{\frac{1}{2}u}^{M}\right)+n\left(E_{\frac{3}{2}u}^{M}\right)\nonumber \\
 & = & n\left(E_{\frac{1}{2}g}^{Z}\right)+n\left(E_{\frac{3}{2}g}^{Z}\right)+n\left(E_{\frac{1}{2}u}^{Z}\right)+n\left(E_{\frac{3}{2}u}^{Z}\right)\nonumber \\
 & = & n\left(E_{\frac{1}{2}g}^{A}\right)+n\left(E_{\frac{3}{2}g}^{A}\right)+n\left(E_{\frac{1}{2}u}^{A}\right)+n\left(E_{\frac{3}{2}u}^{A}\right)\nonumber \\
 & = & n\left(E_{\frac{1}{2}g}^{X}\right)+n\left(E_{\frac{1}{2}u}^{X}\right)\nonumber \\
 & = & n\left(E_{\frac{1}{2}g}^{R}\right)+n\left(E_{\frac{1}{2}u}^{R}\right)
\end{eqnarray}
 and two constraints on the angular momentum conservation along $\Gamma Z$ and $MA$
\begin{equation}
n\left(E_{\frac{1}{2}g}^{\Gamma}\right)+n\left(E_{\frac{1}{2}u}^{\Gamma}\right)=n\left(E_{\frac{1}{2}g}^{Z}\right)+n\left(E_{\frac{1}{2}u}^{Z}\right)
\end{equation}
\begin{equation}
n\left(E_{\frac{1}{2}g}^{M}\right)+n\left(E_{\frac{1}{2}u}^{M}\right)=n\left(E_{\frac{1}{2}g}^{A}\right)+n\left(E_{\frac{1}{2}u}^{A}\right)
\end{equation}
Solve these linear equations, we get 13 independent BR generators, wherein 10 of them are
  atomic BRs, while the other 3  are not.
The three nontrivial generators can be chosen as
\begin{align}
\mathbf{B}_{\mathbb{Z}_2} & =
E_{\frac{3}{2}g}^{\Gamma} - 2E_{\frac{1}{2}g}^{\Gamma} - E_{\frac{3}{2}u}^{\Gamma} + 2E_{\frac{1}{2}u}^{\Gamma} \nonumber \\
& - E_{\frac{3}{2}g}^{M} + E_{\frac{3}{2}u}^{M} -6 E_{\frac{3}{2}g}^Z + 6E_{\frac{3}{2}u}^Z
\end{align}
\begin{equation}
\mathbf{B}_{\mathbb{Z}_4} =
  E_{\frac{3}{2}g}^{\Gamma} - E_{\frac{3}{2}u}^{\Gamma}
- E_{\frac{3}{2}g}^{Z} + E_{\frac{3}{2}u}^{Z}
\end{equation}
\begin{equation}
\mathbf{B}_{\mathbb{Z}_8} =
E_{\frac{3}{2}g}^{\Gamma} - E_{\frac{3}{2}u}^{\Gamma}
\end{equation}
Here $\mathbb{Z}_N$ in the subscript represents that $N\mathbf{B}_{\mathbb{Z}_N}$ is an atomic BR.
Therefore, the BRs of $P4/m$ can be classified into $2\times 4\times 8$ classes, each of them is
 given by three integers $(mnl)$ that define the representative BR
 $m\mathbf{B}_{\mathbb{Z}_2} + n\mathbf{B}_{\mathbb{Z}_4} + l\mathbf{B}_{\mathbb{Z}_8}$,
 with $m=0,1$, $n=0,1,2,3$, and $l=0,1\cdots 7$.

\subsection{Understand the indicators}\label{sub:UnderstandID}
From the parity criterion, we find that all these three generators correspond to  weak or strong topological insulators.
Specifically, $\mathbf{B}_{\mathbb{Z}_2}$ has a nontrivial weak index $\mathbb{Z}_2(110;0)$,
 $\mathbf{B}_{\mathbb{Z}_4}$ also has a nontrivial weak index  $\mathbb{Z}_2(001;0)$,
 whereas  $\mathbf{B}_{\mathbb{Z}_8}$ has a nontrivial strong index $\mathbb{Z}_2(000;1)$.

The double of $\mathbf{B}_{\mathbb{Z}_4}$ or $\mathbf{B}_{\mathbb{Z}_8}$ corresponds to
 mirror ($M_z=PC_2$) protected topological crystalline insulators,
 wherein $2\mathbf{B}_{\mathbb{Z}_4}$ has mirror Chern number $2$ ($\mathrm{mod}\ 4$)
 in both the $k_z=0$- and $k_z=\pi$-slices,
 while $2\mathbf{B}_{\mathbb{Z}_8}$ has mirror Chern number $2$ ($\mathrm{mod}\ 4$)
 in the $k_z=0$-slice and mirror  Chern number $0$ ($\mathrm{mod}\ 4$) in the $k_z=\pi$-slice.
This can be proved by firstly divide the eigenstates at $k_z=0$-slice ($k_z=\pi$-slice) into
 two sectors according to their $M_z$ eigenvalues and then apply the
 following Chern number Fu-Kane-like formula in each sector \cite{fang_C4}
\begin{equation}\label{mChern}
i^C = (-1)^{N_{occ}}\prod_{n\in \mathrm{occ}} \xi_n (\Gamma) \xi_n(M) \zeta_n (X)
\end{equation}
Here $\xi_n(\Gamma)$ is the $C_4$ eigenvalue of the $n$-band at $\Gamma$,
 $\xi_n(M)$ is the $C_4$ eigenvalue of the $n$-band at $M$,
 $\zeta_n(X)$ is the $C_2$ eigenvalue of the $n$-band at $X$,
 and $C$ is the Chern number.

Now the only left element is $4\mathbf{B}_{\mathbb{Z}_8}$.
To figure out it, let us firstly generalize the mappings in Eq. (\ref{eq:2Dmap-1a12})-(\ref{eq:2Dmap-1b32})
 to the case with inversion symmetry.
Treating $k_z=0$- and $k_z=\pi$-slices as two 2D systems
 and applying Eq. (\ref{eq:Dk-def}), we find that
\begin{equation}
E_{\frac{1}{2}g}^{1a}(0/\pi) \mapsto E_{\frac{1}{2}g}^{\Gamma/Z} + E_{\frac{1}{2}g}^{M/A} + E_{\frac{1}{2}g}^{X/R}
\end{equation}
\begin{equation}
E_{\frac{3}{2}g}^{1a}(0/\pi) \mapsto E_{\frac{3}{2}g}^{\Gamma/Z} + E_{\frac{3}{2}g}^{M/A} + E_{\frac{1}{2}g}^{X/R}
\end{equation}
\begin{equation}
E_{\frac{1}{2}u}^{1a}(0/\pi) \mapsto E_{\frac{1}{2}u}^{\Gamma/Z} + E_{\frac{1}{2}u}^{M/A} + E_{\frac{1}{2}u}^{X/R}
\end{equation}
\begin{equation}
E_{\frac{3}{2}u}^{1a}(0/\pi) \mapsto E_{\frac{3}{2}u}^{\Gamma/Z} + E_{\frac{3}{2}u}^{M/A} + E_{\frac{1}{2}u}^{X/R}
\end{equation}
\begin{equation}
E_{\frac{1}{2}g}^{1b}(0/\pi) \mapsto E_{\frac{1}{2}g}^{\Gamma/Z} + E_{\frac{3}{2}g}^{M/A} + E_{\frac{1}{2}u}^{X/R}
\end{equation}
\begin{equation}
E_{\frac{3}{2}g}^{1b}(0/\pi) \mapsto E_{\frac{3}{2}g}^{\Gamma/Z} + E_{\frac{1}{2}g}^{M/A} + E_{\frac{1}{2}u}^{X/R}
\end{equation}
\begin{equation}
E_{\frac{1}{2}u}^{1b}(0/\pi) \mapsto E_{\frac{1}{2}u}^{\Gamma/Z} + E_{\frac{3}{2}u}^{M/A} + E_{\frac{1}{2}g}^{X/R}
\end{equation}
\begin{equation}
E_{\frac{3}{2}u}^{1b}(0/\pi) \mapsto E_{\frac{3}{2}u}^{\Gamma/Z} + E_{\frac{1}{2}u}^{M/A} + E_{\frac{1}{2}g}^{X/R}
\end{equation}
Therefore, we find that the $4\mathbf{B}_{\mathbb{Z}_8}$ can be interpreted as
\begin{align}
  & E_{\frac{1}{2}g}^{1b}(0) + E_{\frac{3}{2}g}^{1b}(0)
+ E_{\frac{1}{2}g}^{1a}(\pi) + E_{\frac{3}{2}g}^{1a}(\pi) \nonumber \\
\mapsto \quad & 4\mathbf{B}_{\mathbb{Z}_8}  \textrm{ mod an atomic BR}
\end{align}
which implies the WC flow from the plaquette center ($k_z=0$) to the site ($k_z=\pi$).
As all the atomic BRs can not imply any WC flow, all the BRs in the $(004)$ class must have the nontrivial $\mathbb{Z}_2$-flow.

\subsection{Matlab script}
Here we also provide a Matlab script to automatically calculate the symmetry indicator of a given BR of the space group $P4/m$.
\begin{widetext}
\begin{lstlisting}
clear;
%
% bases =========================================
%
e_G12g=zeros(20,1); e_G12g(1)=1; e_G32g=zeros(20,1); e_G32g(2)=1;
e_G12u=zeros(20,1); e_G12u(3)=1; e_G32u=zeros(20,1); e_G32u(4)=1;

e_M12g=zeros(20,1); e_M12g(5)=1; e_M32g=zeros(20,1); e_M32g(6)=1;
e_M12u=zeros(20,1); e_M12u(7)=1; e_M32u=zeros(20,1); e_M32u(8)=1;

e_X12g=zeros(20,1); e_X12g(9)=1; e_X12u=zeros(20,1); e_X12u(10)=1;

e_Z12g=zeros(20,1); e_Z12g(11)=1; e_Z32g=zeros(20,1); e_Z32g(12)=1;
e_Z12u=zeros(20,1); e_Z12u(13)=1; e_Z32u=zeros(20,1); e_Z32u(14)=1;

e_A12g=zeros(20,1); e_A12g(15)=1; e_A32g=zeros(20,1); e_A32g(16)=1;
e_A12u=zeros(20,1); e_A12u(17)=1; e_A32u=zeros(20,1); e_A32u(18)=1;

e_R12g=zeros(20,1); e_R12g(19)=1; e_R12u=zeros(20,1); e_R12u(20)=1;
%
% AI bases ======================================
%
AA=[
e_G12g + e_M12g + e_X12g + e_Z12g + e_A12g + e_R12g, ...     % 1a 1/2g
e_G32g + e_M32g + e_X12g + e_Z32g + e_A32g + e_R12g, ...     % 1a 3/2g
e_G12u + e_M12u + e_X12u + e_Z12u + e_A12u + e_R12u, ...     % 1a 1/2u
e_G32u + e_M32u + e_X12u + e_Z32u + e_A32u + e_R12u, ...     % 1a 3/2u
e_G12g + e_M12g + e_X12g + e_Z12u + e_A12u + e_R12u, ...     % 1b 1/2g
e_G32g + e_M32g + e_X12g + e_Z32u + e_A32u + e_R12u, ...     % 1b 3/2g
e_G12g + e_M32g + e_X12u + e_Z12g + e_A32g + e_R12u, ...     % 1c 1/2g
e_G32g + e_M12g + e_X12u + e_Z32g + e_A12g + e_R12u, ...     % 1c 3/2g
e_G12u + e_M32u + e_X12g + e_Z12u + e_A32u + e_R12g, ...     % 1c 1/2u
e_G12g + e_M32g + e_X12u + e_Z12u + e_A32u + e_R12g, ...     % 1d 1/2g
e_G32g + e_M12g + e_X12u + e_Z32u + e_A12u + e_R12g, ...     % 1d 3/2g
e_G12g + e_G32g + e_M12u + e_M32u + e_X12g + e_X12u ...
  + e_Z12g + e_Z32g + e_A12u + e_A32u + e_R12g + e_R12u, ... % 2e 1/2g
e_G12g + e_G32g + e_M12u + e_M32u + e_X12g + e_X12u ...
  + e_Z12u + e_Z32u + e_A12g + e_A32g + e_R12g + e_R12u      % 2f 1/2g
];
%
% Nontrivial generators =========================
%
BZ2 = e_G32g - 2*e_G12g - e_G32u + 2*e_G12u ...
    - e_M32g + e_M32u - 6*e_Z32g + 6*e_Z32u;
BZ4 = e_G32g - e_G32u - e_Z32g + e_Z32u;
BZ8 = e_G12u - e_G12g;
%
%  diagnose the BR ==============================
%
%BR = e_G12g + e_G32u + e_M12u + e_M32g + e_X12g + e_X12u ...
%   + e_Z12u + e_Z32g + e_A12u + e_A32g + e_R12g + e_R12u;
BR = e_G12g + e_G32g + e_M12g + e_M32g + e_X12g + e_X12g ...
   + e_Z12g + e_Z32g + e_A12g + e_A32g + e_R12u + e_R12u;
%
[m, n, l] = fun_class83( BR, AA, BZ2, BZ4, BZ8);
fprintf('m, n, l= %d, %d, %d\n',m,n,l);
%
% subroutine to calculate the indicator =========
%
function [ n1, n2, n3 ] = fun_class83( bb, AA, b1,b2,b3 )
tol=1e-3;
for n1=0:1
    for n2=0:3
        for n3=0:7
            btmp = n1*b1+n2*b2+n3*b3;
            CC=AA\(btmp - bb);
            err = norm(CC-round(CC));
            %
            %fprintf('%d %d %d %f\n', n1, n2, n3, err)
            %
            if err < tol
                break;
            end
        end
        %
        if err < tol
            break;
        end
    end
    %
    if err < tol
        break;
    end
end

if err>=tol
    n1=-1;
    n2=-1;
    n3=-1;
end

end
\end{lstlisting}

\end{widetext}

where the variable BR is the input band representation,
and $(mnl)$ is the indicator calculated from BR.
For practical application, one only need to replace the definition of BR (line $52$ in the code).
It should be noticed that, this script is not limited to the space group $P4/m$, but is applicable to any space group containing $C_4$ and inversion symmetries.

We have used the occupied BR of our 3D model (the commented definition at line $50$ in the code) to verify the symmetry indicator, which indeed gives $(004)$.

\subsection{Fu-Kane formula}

\begin{table*}
\begin{tabular}{|c|c|c|c|}
\hline
Lattice & Space groups & $n$ & Definitions for $n_{\frac{3}{2}}^{+}$, $n_{\frac{3}{2}}^{-}$, $n_{\frac{1}{2}}^{+}$,
$n_{\frac{1}{2}}^{-}$\tabularnewline
\hline
\hline
\multirow{4}{*}{$\Gamma_{q}$} & \multirow{4}{*}{83, 123, 124, 127, 128} & $n_{\frac{1}{2}}^{+}$ & $n(E_{\frac{1}{2}g}^{\Gamma})+n(E_{\frac{1}{2}g}^{M})+n(E_{\frac{1}{2}g}^{Z})+n(E_{\frac{1}{2}g}^{A})+n(E_{\frac{1}{2}g}^{X})+n(E_{\frac{1}{2}g}^{R})$\tabularnewline
\cline{3-4}
 &  & $n_{\frac{1}{2}}^{-}$ & $n(E_{\frac{1}{2}u}^{\Gamma})+n(E_{\frac{1}{2}u}^{M})+n(E_{\frac{1}{2}u}^{Z})+n(E_{\frac{1}{2}u}^{A})+n(E_{\frac{1}{2}u}^{X})+n(E_{\frac{1}{2}u}^{R})$\tabularnewline
\cline{3-4}
 &  & $n_{\frac{3}{2}}^{+}$ & $n(E_{\frac{3}{2}g}^{\Gamma})+n(E_{\frac{3}{2}g}^{M})+n(E_{\frac{3}{2}g}^{Z})+n(E_{\frac{3}{2}g}^{A})+n(E_{\frac{1}{2}g}^{X})+n(E_{\frac{1}{2}g}^{R})$\tabularnewline
\cline{3-4}
 &  & $n_{\frac{3}{2}}^{-}$ & $n(E_{\frac{3}{2}u}^{\Gamma})+n(E_{\frac{3}{2}u}^{M})+n(E_{\frac{3}{2}u}^{Z})+n(E_{\frac{3}{2}u}^{A})+n(E_{\frac{1}{2}u}^{X})+n(E_{\frac{1}{2}u}^{R})$\tabularnewline
\hline
\multirow{4}{*}{$\Gamma_{q}^{v}$} & \multirow{4}{*}{87, 139, 140} & $n_{\frac{1}{2}}^{+}$ & $n(E_{\frac{1}{2}g}^{\Gamma})+n(E_{\frac{1}{2}g}^{M})+n(E_{\frac{1}{2}g}^{X})+2n(E_{\frac{1}{2}g}^{N})+n\left(E_{\frac{1}{2}}^{P}\right)$\footnotemark[1]\tabularnewline
\cline{3-4}
 &  & $n_{\frac{1}{2}}^{-}$ & $n(E_{\frac{1}{2}u}^{\Gamma})+n(E_{\frac{1}{2}u}^{M})+n(E_{\frac{1}{2}u}^{X})+2n(E_{\frac{1}{2}u}^{N})+n\left(E_{\frac{3}{2}}^{P}\right)$\tabularnewline
\cline{3-4}
 &  & $n_{\frac{3}{2}}^{+}$ & $n(E_{\frac{3}{2}g}^{\Gamma})+n(E_{\frac{3}{2}g}^{M})+n(E_{\frac{1}{2}g}^{X})+2n(E_{\frac{1}{2}g}^{N})+n\left(E_{\frac{3}{2}}^{P}\right)$\tabularnewline
\cline{3-4}
 &  & $n_{\frac{3}{2}}^{-}$ & $n(E_{\frac{3}{2}u}^{\Gamma})+n(E_{\frac{3}{2}u}^{M})+n(E_{\frac{1}{2}u}^{X})+2n(E_{\frac{1}{2}u}^{N})+n\left(E_{\frac{1}{2}}^{P}\right)$\tabularnewline
\hline
\multirow{4}{*}{$\Gamma_{c}$} & \multirow{4}{*}{221} & $n_{\frac{1}{2}}^{+}$ & $n(E_{\frac{1}{2}g}^{\Gamma})+n(F_{\frac{3}{2}g}^{\Gamma})+n(E_{\frac{1}{2}g}^{R})+n(F_{\frac{3}{2}g}^{R})+2n(E_{\frac{1}{2}g}^{M})+n(E_{\frac{3}{2}g}^{M})+2n(E_{\frac{1}{2}g}^{X})+n(E_{\frac{3}{2}g}^{X})$\tabularnewline
\cline{3-4}
 &  & $n_{\frac{1}{2}}^{-}$ & $n(E_{\frac{1}{2}u}^{\Gamma})+n(F_{\frac{3}{2}u}^{\Gamma})+n(E_{\frac{1}{2}u}^{R})+n(F_{\frac{3}{2}u}^{R})+2n(E_{\frac{1}{2}u}^{M})+n(E_{\frac{3}{2}u}^{M})+2n(E_{\frac{1}{2}u}^{X})+n(E_{\frac{3}{2}u}^{X})$\tabularnewline
\cline{3-4}
 &  & $n_{\frac{3}{2}}^{+}$ & $n(F_{\frac{3}{2}g}^{\Gamma})+n(E_{\frac{5}{2}g}^{\Gamma})+n(F_{\frac{3}{2}g}^{R})+n(E_{\frac{5}{2}g}^{R})+2n(E_{\frac{3}{2}g}^{M})+n(E_{\frac{1}{2}g}^{M})+2n(E_{\frac{3}{2}g}^{X})+n(E_{\frac{1}{2}g}^{X})$\tabularnewline
\cline{3-4}
 &  & $n_{\frac{3}{2}}^{-}$ & $n(F_{\frac{3}{2}u}^{\Gamma})+n(E_{\frac{5}{2}u}^{\Gamma})+n(F_{\frac{3}{2}u}^{R})+n(E_{\frac{5}{2}u}^{R})+2n(E_{\frac{3}{2}u}^{M})+n(E_{\frac{1}{2}u}^{M})+2n(E_{\frac{3}{2}u}^{X})+n(E_{\frac{1}{2}u}^{X})$\tabularnewline
\hline
\multirow{8}{*}{$\Gamma_{c}^{f}$} & \multirow{4}{*}{225} & $n_{\frac{1}{2}}^{+}$ & $n(E_{\frac{1}{2}g}^{\Gamma})+n(F_{\frac{3}{2}g}^{\Gamma})+2n(E_{\frac{1}{2}g}^{X})+n(E_{\frac{3}{2}g}^{X})+2n(E_{\frac{1}{2}g}^{L})+2n(E_{\frac{3}{2}g}^{L})+n(E_{\frac{1}{2}}^{W})$\tabularnewline
\cline{3-4}
 &  & $n_{\frac{1}{2}}^{-}$ & $n(E_{\frac{1}{2}u}^{\Gamma})+n(F_{\frac{3}{2}u}^{\Gamma})+2n(E_{\frac{1}{2}u}^{X})+n(E_{\frac{3}{2}u}^{X})+2n(E_{\frac{1}{2}u}^{L})+2n(E_{\frac{3}{2}u}^{L})+n(E_{\frac{3}{2}}^{W})$\tabularnewline
\cline{3-4}
 &  & $n_{\frac{3}{2}}^{+}$ & $n(F_{\frac{3}{2}g}^{\Gamma})+n(E_{\frac{5}{2}g}^{\Gamma})+2n(E_{\frac{3}{2}g}^{X})+n(E_{\frac{1}{2}g}^{X})+2n(E_{\frac{1}{2}g}^{L})+2n(E_{\frac{3}{2}g}^{L})+n(E_{\frac{3}{2}}^{W})$\tabularnewline
\cline{3-4}
 &  & $n_{\frac{3}{2}}^{-}$ & $n(F_{\frac{3}{2}u}^{\Gamma})+n(E_{\frac{5}{2}u}^{\Gamma})+2n(E_{\frac{3}{2}u}^{X})+n(E_{\frac{1}{2}u}^{X})+2n(E_{\frac{1}{2}u}^{L})+2n(E_{\frac{3}{2}u}^{L})+n(E_{\frac{1}{2}}^{W})$\tabularnewline
\cline{2-4}
 & \multirow{4}{*}{226} & $n_{\frac{1}{2}}^{+}$ & $n(E_{\frac{1}{2}g}^{\Gamma})+n(F_{\frac{3}{2}g}^{\Gamma})+n(E_{\frac{3}{2}g}^{X})+n(E_{\frac{1}{2}u}^{X})+n(E_{\frac{3}{2}u}^{X})$\tabularnewline
\cline{3-4}
 &  & $n_{\frac{1}{2}}^{-}$ & $n(E_{\frac{1}{2}u}^{\Gamma})+n(F_{\frac{3}{2}u}^{\Gamma})+n(E_{\frac{3}{2}u}^{X})+n(E_{\frac{1}{2}g}^{X})+n(E_{\frac{3}{2}g}^{X})$\tabularnewline
\cline{3-4}
 &  & $n_{\frac{3}{2}}^{+}$ & $n(F_{\frac{3}{2}g}^{\Gamma})+n(E_{\frac{5}{2}g}^{\Gamma})+n(E_{\frac{1}{2}g}^{X})+n(E_{\frac{1}{2}u}^{X})+n(E_{\frac{3}{2}u}^{X})$\tabularnewline
\cline{3-4}
 &  & $n_{\frac{3}{2}}^{-}$ & $n(F_{\frac{3}{2}u}^{\Gamma})+n(E_{\frac{5}{2}u}^{\Gamma})+n(E_{\frac{1}{2}u}^{X})+n(E_{\frac{1}{2}g}^{X})+n(E_{\frac{3}{2}g}^{X})$\tabularnewline
\hline
\multirow{4}{*}{$\Gamma_{c}^{v}$} & \multirow{4}{*}{229} & $n_{\frac{1}{2}}^{+}$ & $n(E_{\frac{1}{2}g}^{\Gamma})+n(F_{\frac{3}{2}g}^{\Gamma})+n(E_{\frac{1}{2}g}^{H})+n(F_{\frac{3}{2}g}^{H})+3n(E_{\frac{1}{2}g}^{N})+n\left(E_{\frac{1}{2}}^{P}\right)+n\left(F_{\frac{3}{2}}^{P}\right)$\tabularnewline
\cline{3-4}
 &  & $n_{\frac{1}{2}}^{-}$ & $n(E_{\frac{1}{2}u}^{\Gamma})+n(F_{\frac{3}{2}u}^{\Gamma})+n(E_{\frac{1}{2}u}^{H})+n(F_{\frac{3}{2}u}^{H})+3n(E_{\frac{1}{2}u}^{N})+n\left(F_{\frac{3}{2}}^{P}\right)+n\left(E_{\frac{5}{2}}^{P}\right)$\tabularnewline
\cline{3-4}
 &  & $n_{\frac{3}{2}}^{+}$ & $n(F_{\frac{3}{2}g}^{\Gamma})+n(E_{\frac{5}{2}g}^{\Gamma})+n(F_{\frac{3}{2}g}^{H})+n(E_{\frac{5}{2}g}^{H})+3n(E_{\frac{1}{2}g}^{N})+n\left(F_{\frac{3}{2}}^{P}\right)+n\left(E_{\frac{5}{2}}^{P}\right)$\tabularnewline
\cline{3-4}
 &  & $n_{\frac{3}{2}}^{-}$ & $n(F_{\frac{3}{2}u}^{\Gamma})+n(E_{\frac{5}{2}u}^{\Gamma})+n(F_{\frac{3}{2}u}^{H})+n(E_{\frac{5}{2}u}^{H})+3n(E_{\frac{1}{2}u}^{N})+n\left(E_{\frac{1}{2}}^{P}\right)+n\left(F_{\frac{3}{2}}^{P}\right)$\tabularnewline
\hline
\end{tabular}

\footnotetext[1]{In space group 87, the little group at $N$ is $C_i$ and the corresponding irreps are denoted as $A_{\frac{1}{2}g}$ and  $A_{\frac{1}{2}u}$ in Ref. [\onlinecite{point-group}], both of which are one dimensional. However, due to the Kramers theorem, the irreps at $N$ should be double degenerate, thus we adopt the two dimensional notations $E_{\frac{1}{2}g}$ and  $E_{\frac{1}{2}u}$.}

\protect\caption{ The concrete expressions for $n_{\frac{3}{2}}^{+}$, $n_{\frac{3}{2}}^{-}$, $n_{\frac{1}{2}}^{+}$, $n_{\frac{1}{2}}^{-}$
in the $\mathbb{Z}_{8}$ Fu-Kane-like formulas in all applicable space groups. The notations
of high symmetry momenta follow Ref. [\onlinecite{Bilbao_BZ_2014}],
and the notations of point group irreps follow Ref. [\onlinecite{point-group}]. }

\label{tab-Z8}

\end{table*}

As discussed in section \ref{sub:UnderstandID}, the $\mathbb{Z}_2$ indicator $m$ and the $\mathbb{Z}_4$ indicator $n$ can be indeed calculated from the knowledge of inversion and rotation eigenvalues.
Specifically, we have
\begin{equation}
(-1)^m = \prod_{\mathbf{K}} \prod_{n\in \mathrm{occ}} \lambda_n(\mathbf{K})
\end{equation}
\begin{equation}
i^n = i^{C_\mathrm{m}(\pi)}
\end{equation}
where $\lambda_n(\mathbf{K})$ is the parity of the $n$-th Krammer pair at the momentum $\mathbf{K}$, $\mathbf{K}$ goes over the four time reversal invariant momenta (TRIM) with $k_x=\pi$, and $C_m(\pi)$ is the mirror Chern number of $k_z=\pi$ slice (mod 4) which can be calculated from Eq. (\ref{mChern}).
Here we also find a similar formula to calculate the $\mathbb{Z}_8$ indicator $l$ from the symmetry eigenvalues directly
\begin{equation}
l = \left( n_\frac{1}{2}^- + 3n_\frac{3}{2}^+ - n_\frac{1}{2}^+ - 3n_\frac{3}{2}^- \right) /2 \mod 8
\end{equation}
which is applicable to all space groups with inversion and symmorphic $C_4$ rotation.
The concrete definition of $n_\frac{1}{2}^+$, $n_\frac{1}{2}^-$, $n_\frac{3}{2}^+$, $n_\frac{3}{2}^-$
in various groups are listed in table \ref{tab-Z8}.

\end{document}